\documentclass[a4paper,fleqn]{cas-dc}

 \usepackage[numbers]{natbib}

\usepackage{tikz}
\usepackage{blindtext}
\usepackage{adjustbox}
\usepackage{makecell}
\usepackage{needspace}
\usepackage{afterpage}  
\usepackage{caption}
\usepackage{float}
\usepackage{placeins}
\usepackage{multicol}

\def\tsc#1{\csdef{#1}{\textsc{\lowercase{#1}}\xspace}}
\tsc{WGM}
\tsc{QE}
\tsc{EP}
\tsc{PMS}
\tsc{BEC}
\tsc{DE}

\begin{document}



\title [mode = title]{Self-supervised learning for skin cancer diagnosis with limited training data}


 

 \author[3]{Hamish Haggerty   \corref{cor1}} 
 \ead{hamish.haggerty@uqconnect.edu.au}

  \author[3]{Rohitash Chandra  \corref{cor2}} 
  \ead{rohitash.chandra@unsw.edu.au} 

  \affiliation[3]{Transitional Artificial Intelligence Research Group, School of Mathematics and Statistics, UNSW Sydney, Sydney. Australia}

 \cortext[cor1]{Corresponding author}
  \cortext[cor2]{Principal corresponding author}



 \begin{abstract} 
Early cancer detection is crucial for prognosis, but many cancer types lack large labelled datasets required for developing deep learning models. This paper investigates self-supervised learning (SSL) as an alternative to the standard supervised pre-training on ImageNet for scenarios with limited training data using a deep learning model (ResNet-50). We first demonstrate that SSL pre-training on ImageNet (via the Barlow Twins SSL algorithm) outperforms supervised pre-training (SL) using a skin lesion dataset with limited training samples. We then consider \textit{further} SSL pre-training (of the two ImageNet pre-trained models) on task-specific datasets, where our implementation is motivated by supervised transfer learning. This approach significantly enhances initially SL pre-trained models, closing the performance gap with initially SSL pre-trained ones. Surprisingly, further pre-training on just the limited fine-tuning data achieves this performance equivalence. 
Linear probe experiments reveal that improvement stems from enhanced feature extraction.  Hence, we find that minimal further SSL pre-training on task-specific data can be as effective as large-scale SSL pre-training on ImageNet for medical image classification tasks with limited labelled data. 
We validate these results on an oral cancer histopathology dataset, suggesting broader applicability across medical imaging domains facing labelled data scarcity.

\end{abstract}



\maketitle 
\section{Introduction}

Cancer is the second leading cause of death worldwide, with almost 10 million deaths estimated in 2020 \cite{ferlay2021cancer}. In several types of cancer (e.g. skin, oral, pancreatic), early diagnosis is the major determining factor in prognosis \cite{jeschke2012biomarkers,filella2016prostate,kazarian2017testing}. If cancer is detected sufficiently early, survival rates may be above 90\% \cite{rigel2010evolution,kamisawa2016pancreatic,warnakulasuriya2009global}. On the other hand, the prevalence of several cancer types is increasing \cite{warnakulasuriya2009global}, and this is particularly true in poorer communities in the developing world \cite{song2021classification}. This is of particular concern since individuals in such communities have lower access to the expert clinicians traditionally needed to diagnose such cancer. There is thus an urgent need to develop cheaper and more data-efficient methods of diagnosis that can be deployed globally. In this context, machine learning models that can effectively utilise limited datasets are particularly crucial, as they could enable the development of artificial intelligence-assisted diagnostic tools \cite{
kaur2020medical,syed2020artificial,szolovits1988artificial} in areas where large, labelled datasets are not available. 


In recent years, there has been significant interest in applying machine learning to cancer image diagnosis \cite{Litjens2017}. This includes lung, breast and several other forms of cancer, and involves classifying clinical images into categories (`malignant' or `benign') or more fine-grained classification \cite{kwong2021survey,dildar2021skin,Yadav2019}.
There have been some challenges in the use of deep learning models, since they require large and class-balanced datasets to function properly. The availability of timely data with proper organisation has been a challenge in the medical domain due to restrictions in archive access given patient confidentiality and ethical approval concerns \cite{thapa2021precision}.  
In general, assembling labelled datasets is a complex and resource-intensive process regarding collection and annotation \cite{Deng2009}. This is even more pronounced for medical images, which require expert knowledge or even medical testing (e.g. via biopsy) to classify, dramatically increasing the cost \cite{Armato2011,Clark2013}. Therefore, for several types of cancer, we lack image-based datasets that are suitable for training machine learning models.

Sengupta et al. \cite{sengupta2022scarcity} highlighted the scarcity of publicly available oral cancer datasets. Although there has been research published on oral cancer image classification, the data used is closed source. Our attempt to obtain data by contacting primary authors was not successful, which further highlights the problem of data availability and unreproducible research \cite{jarvis2016irreproducibility,shiffrin2018scientific}. Oral cancer is diagnosed through biopsy, but the decision to biopsy is made by visual inspection of the mouth \cite{baykul2010early}, and early detection is crucial for prognosis. In the United States, the 5-year survival with early detection is 85\%, but only 28\% of cases are detected at this stage. In later stages, when cancer has spread, 5-year survival drops to 40\% \cite{cancer_net_2022}. The lack of available datasets is concerning given that the prevalence of oral cancer is increasing, \cite{warnakulasuriya2009global}, particularly among poorer parts of the developing world, where expert medical care is less available. As an additional point, certain kinds of cancer have extremely low prevalence. For example, Chordoma has a prevalence of only 0.18-0.84 per million \cite{Bakker2018}, so constructing a large labelled dataset in such cases is very difficult.  Developing data-efficient techniques is therefore essential since factors such as disease rarity, privacy concerns, and labelling costs result in the absence of labelled datasets for several cancer types. In this study, we address this challenge by investigating self-supervised learning approaches and propose an efficient technique for enhancing supervised pre-trained models.

Supervised training of deep learning models involves initialising the weights of the network, classically via random initialisation, and then iteratively updating the weights through variants of the backpropagation algorithm \cite{goodfellow2016deep}. A  well-known problem with random weight initialisation is the requirement for large amounts of labelled data to achieve good performance. Transfer learning is a machine learning technique that attempts to address this problem by instead initialising the weights with those of a network that has already been trained on a related task - typically through supervised learning \cite{zhuang2020comprehensive}. A common strategy in computer vision applications is to use supervised pre-training on ImageNet \cite{zhuang2020comprehensive}, a large-scale dataset of natural images spanning a diverse range of object categories. It has been demonstrated that transfer learning can lead to significant performance gains compared to training from random initialisation \cite{tan2018survey}.

Self-supervised learning (SSL) \cite{jaiswal2020survey} is an alternative approach to pre-training models, also known as self-supervised pre-training. Unlike supervised pre-training, which requires labelled datasets, SSL utilises \textit{unlabelled} data. This data is used to create a `surrogate task' or `supervisory signal' for learning. For example, SSL has achieved remarkable success in  NLP through the pre-training of large language models \cite{vaswani2017attention,minaee2024largelanguagemodelssurvey}, where the task is typically to predict the next word (or token). It has been more challenging to apply SSL to computer vision tasks since visual images are inherently of higher dimension \cite{lecun2021self}, and developing surrogate tasks has been more difficult than in NLP. However, recently SSL has had success in computer vision tasks as well \cite{jaiswal2020survey,lecun2021self}, and may be particularly applicable in medical image domains where collection of labelled data poses challenges.

The use of transfer learning is common in the medical diagnosis literature for image data \cite{zhuang2020comprehensive}, such as skin cancer \cite{kim2022transfer} and lung cancer diagnosis \cite{Wang2020}. Usually, the models involved have been pre-trained on ImageNet in a supervised fashion. Recent work has also explored the use of SSL pre-training \cite{Huang2023}; however, in the low data regime, it is still unclear which pre-training methodology is superior. In this study, we investigate the efficacy of different pre-training strategies for the downstream task of cancer diagnosis in limited-data scenarios, mainly focusing on skin lesion classification as a model. Specifically:
\begin{enumerate}

\item We compare the standard supervised pre-training on ImageNet with self-supervised pre-training, (via Barlow Twins) where both networks are fine-tuned on a small labelled skin lesion dataset \cite{andrewmvd2019isic}, with limited training samples compared to test samples. 
   
\item We investigate \textit{further} SSL pre-training of each of the ImageNet pre-trained models on \textit{unlabelled} data that is from a similar distribution to the data used for fine-tuning. These doubly pre-trained networks are fine-tuned and evaluated in the same manner as the initial (pre-trained once) networks.
\end{enumerate}

\section{Related work}

We review the deep learning literature on medical image classification, and then give an overview of self-supervised learning and Barlow Twins.

\subsection{Skin cancer  detection}
\label{sec:cancer_lit} 

Conventional deep learning models using supervised learning have had enormous success in skin lesion classification \cite{dildar2021skin}, which has been facilitated by large curated datasets, such as the International Skin Imaging Collaboration (ISIC) database \cite{ISIC2023}, and other large datasets. The ISIC2018 training dataset is a subset of ISIC which features 10,015 image samples, distributed among 7 categories \cite{isic2018}. Guo and Ashour \cite{guo2018multiple} proposed a Multiple Convolutional Neural Network (MCNN) model for skin lesion classification using this dataset.   The training strategy focuses each subsequent CNN on samples poorly classified by previous models, aiming to improve overall performance on challenging cases. Majtner et al. \cite{majtner2018ensemble} fine-tuned two models on 10,015 labelled ISIC samples using deep learning (VGG16 and GoogLeNet), both of which had been pre-trained via supervised learning on ImageNet, and found the ensemble accuracy was higher than individual accuracy. Maron et al. \cite{MARON201957} utilised  11,444 ISIC skin images using a ResNet-50 pre-trained on ImageNet for the classification of benign and malignant samples that outperformed Dermatologists.

Esteva et al. \cite{esteva2017dermatologist} trained an Inception-v3 Convolutional Neural Network (CNN) on a large dataset of 129,450 lesion images with 2,032 disease categories (i.e. instead of a label of `melanoma' there is a more fine-grained subclassification of amelanotic melanoma, lentigo melanoma etc) achieving performance on-par with Dermatologists.  Brinker et al. \cite{brinker2019deep} considered a balanced training set of 4,204 melanoma and nevi mole images, where the classification had been determined by biopsy. While this training set is smaller than in other studies, note that it involves a binary classification problem, resulting in a larger number of samples per class. They also utilised a transfer learning approach, via a ResNet-50 pre-trained on ImageNet with supervised learning. On the metrics of sensitivity and specificity, they found that the trained network outperformed Dermatologists. 


Our review of skin cancer studies has shown that all the mentioned studies  use supervised pre-training on ImageNet and the training sets used are relatively large to achieve high performance accuracy. The high performance is made possible by the existence of large curated datasets, such as ISIC. In our study, we consider skin lesion classification on a much smaller dataset to demonstrate model development for situations where large labelled datasets do not exist, such as oral cancer or rare cancers. We also propose a technique that can enhance the fine-tuning performance of supervised pre-trained models. 

\subsubsection{Related  medical diagnosis problems}

Song et al. \cite{song2021classification} studied the problem of oral cancer image classification on an imbalanced training dataset with a CNN pre-trained on ImageNet with supervised learning. They reported that oversampling of minority class during training can effectively address the problem of imbalanced and limited data in this context. The utilisation of ImageNet supervised pre-training using deep learning models has also been applied to 
Alzheimer's disease multiclass classification from MRI images \cite{khan2023transfer}, pancreatic cancer diagnosis \cite{baldota2021deep}, diabetic retinopathy classification from images \cite{jabbar2022transfer}, and other areas of medicine. In other words, supervised pre-training on ImageNet, typically employing CNN-based architectures, is a ubiquitous transfer learning technique not just for cancer image classification, but across various areas of medical image classification. Recent work has also explored using SSL pre-trained models for medical image diagnosis \cite{Huang2023} that is promising in leveraging large amounts of unlabelled data. However, the comparative effectiveness of SSL versus traditional supervised pre-training in medical imaging tasks with limited labelled data remains an open question that our study aims to address.

\subsection{Self-supervised learning}
\begin{figure}[htbp!]
    \centering
    \includegraphics[scale=0.9]{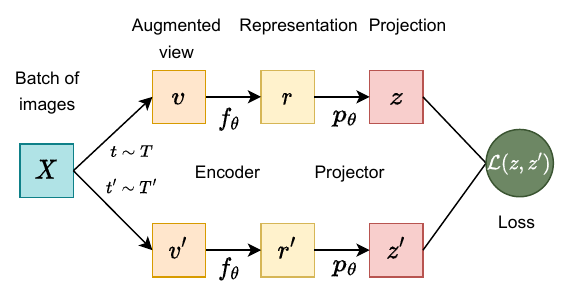}
    \caption{Overview of a joint embedding architecture for SSL. The encoder $f_\theta$ is typically a deep neural network such as a CNN, and the projector $p_\theta$ is typically a feedforward network with several layers. $T$ and $T'$ are distributions of data augmentations. For example, some random amount of cropping, blur etc may be applied with an example using the ISIC2019 data in Figure \ref{fig:isic_bt_cancer}.}
    \label{fig:jea}
\end{figure}

Self-supervised learning (SSL) is a method of pre-training models using unlabelled data. The goal is still to learn an initialisation of a deep network $f_{\theta}$, similar to standard supervised pre-training. One approach to SSL is the joint embedding architecture (JEA) \cite{lecun2021self}, illustrated in Figure \ref{fig:jea}. The intuition of JEA  is that random distortions (e.g., blur) to an image $x$, does not change its semantic content. Suppose $x$ and $y$ are distinct images and $(b(x),b'(x))$ are blurry versions of $x$ - called \textit{positive samples} since they arise from the same image. Conversely, $(b(x),y)$ is a \textit{negative sample}. Then  the network should preserve semantic structure: $f_\theta(b(x)) \approx f_\theta(b'(x)) \approx f_\theta(x) $. However, it's crucial to avoid the trivial solution of collapse to a constant function, which also satisfies this relation. Hence, an additional requirement is that $f_\theta$ is non-trivial; or that distinct images produce distinct representations: $f_\theta(x) \neq f_\theta(y)$. There are two main approaches to prevent this collapse. The contrastive approach involves an explicit comparison between negative samples, with an example algorithm being SimCLR \cite{chen2020simple} (Simple Framework for Contrastive Learning of Visual Representations). The non-contrastive 
approach avoids collapse implicitly through regularisation on the outputs, without requiring comparison via negative samples. Barlow Twins \cite{zbontar2021barlow} is an example of this approach, and we focus on the non-contrastive approach in this study.

The JEA (as in Figure \ref{fig:jea}) requires designated networks $f_{\theta}$ (encoder) and $p_{\theta}$ (projector), a loss function $\mathcal{L}$, and an unlabelled image dataset. Two distributions of data augmentations $T$ and $T'$ are also required. The training process involves sampling a batch of data and computing two distorted views: $v=t(X)$, $v'=t'(X)$, where $t \sim T$ and $t' \sim T'$ are random data augmentations (e.g., cropping, blur). These distorted batches are then passed through the encoder and projector: $z=p_\theta(f_\theta(v))$ and $z'=p_\theta(f_\theta(v'))$. Finally the loss function is computed $\mathcal{L}(z,z')$, which is used to compute gradients for backpropagation. We use the following terminology: the "representation" refers to the output of the encoder: $f_\theta(x)$, the "projected representation" or "projection" is the output of the projector: $p_\theta(f_\theta(x))$. On downstream tasks, these "representations" can be considered with the parameters frozen or unfrozen, i.e. full network fine-tuning or linear probing.

The loss function $\mathcal{L}$ imposes a similarity constraint on $z$ and $z'$, which arise from the same batch of data under different augmentations. Barlow Twins \cite{zbontar2021barlow} is one method of specifying this loss function in SSL. After training, we remove the projector and retain the encoder as the pre-trained model.

One might wonder about the role of the projector $p_\theta$ in Figure \ref{fig:jea}; since only the encoder $f_\theta$ is retained for downstream tasks - why is it necessary to include $p_\theta$? Historically, the loss function was computed in encoder space until Chen et al. \cite{chen2020simple} found that adding a projector to the end of the network during training led to superior representations.  The superior performance also arises when the projector has lower dimensionality than the encoder \cite{zbontar2021barlow} for most algorithms based on the joint embedding architecture. Barlow Twins is an exception, and benefits from having an extremely high projector dimension, as presented in the following section. 
 
\section{Methodology}




\subsection{Barlow Twins}
\label{sec:ssl}

Barlow Twins is an SSL algorithm based on the general joint embedding framework we have outlined. It provides a way of specifying the loss function: $\mathcal{L}(z,z')$. At a high level, the computation of the loss function is extremely simple. The two branch outputs $z$ and $z'$ are normalised $z \leftarrow \text{norm}(z), z' \leftarrow \text{norm}(z')$,  and then multiplied together, yielding a cross correlation matrix $\mathcal{C}=\frac{1}{n}(z')^{\top} z$, which is then equated to the identity matrix  $\mathcal{C}\leftrightarrow I$.

In more detail, if the batch size is $n$ and the projector dimension is $d$, then $z$ and $z'$ will each be $n \times d$.  We first normalise the branches $z$ and $z'$   along the batch dimension and compute the mean, (for the first branch)   $\mu_i = \frac{1}{n} \sum_{k=1}^{n} z_{ki}$ and  the standard deviation: $\sigma_i = \sqrt{\frac{1}{n} \sum_{k=1}^{n} \left(Z_{ki} - \mu_i^A\right)^2}$. We then compute the normalisation as, $z_{ji} \leftarrow \frac{z_{ji}-\mu_i}{\sigma_i}$ which is performed analogously for $z'$. Next, we compute the cross-correlation between the branches with a simple matrix multiplication: $\mathcal{C} = \frac{1}{n}(z')^{\top} z.$ The Barlow Twins loss function then sets $\mathcal{C}$ to the identity matrix: 
\begin{equation}
\mathcal{L_{BT}} = \sum_{i} (\mathcal{C}_{ii}-1)^2 + \lambda  \sum_{i\neq j} \mathcal{C}_{ij}^2
\label{eq:bt}
\end{equation}

where $\lambda$ is a trade-off hyperparameter. One can think of the loss function as being composed of an invariance term and a redundancy reduction term. The invariance term trains the network to ignore the image distortions, and the redundancy reduction term ensures that this is done in a non-trivial way, such that the components of an output vector contain non-redundant information. The hyperparameter  $\lambda$ balances these two objectives.

\begin{figure}[htbp!]
    \centering
    \includegraphics[scale=0.5]{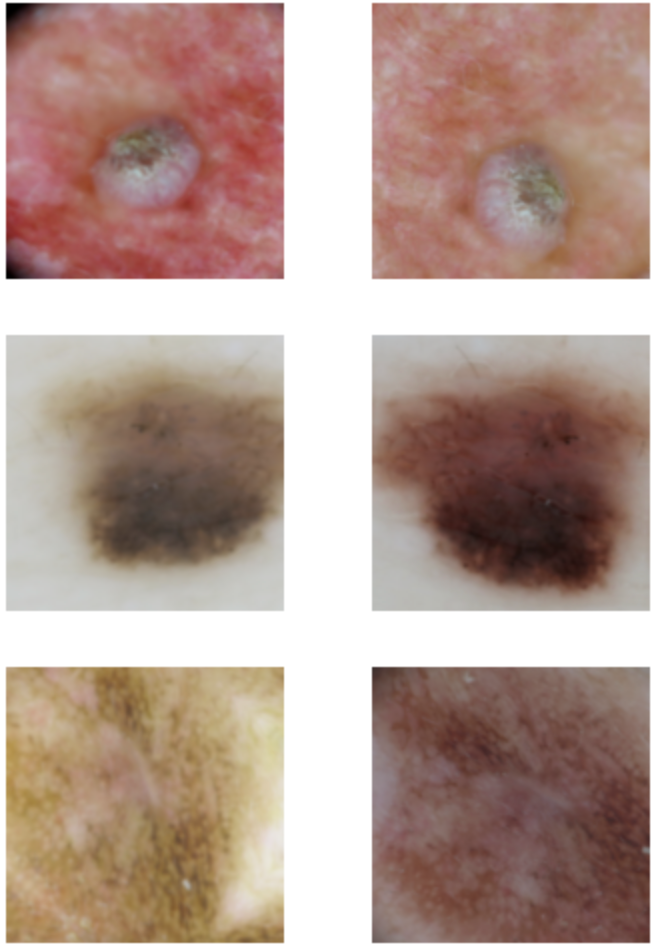}
    \caption{Barlow Twins data augmentation using the ISIC2019 data where we train the network to ignore the row-wise image distortions in a non-redundant way.}
    \label{fig:isic_bt_cancer}
\end{figure}
Next we describe a simple modification of SSL and Barlow Twins framework that  we use in the current study. Specifically, recent work \cite{song2023towards,hua2021feature,jing2022understanding} has found that SSL methods, including Barlow Twins, can suffer from dimensional collapse, where learned embeddings span a low-dimensional space. The projector head partially ameliorates this problem, but it cannot be entirely avoided. Song et al.  \cite{song2023towards} proposed imposing a sparsity constraint on the projector to improve generalisation and reported  downstream performance gains. The method  takes the $L_{2,1}$ norm of the weight matrix of the final layer  $\| W \|_{2,1}$, which is scaled by a hyperparameter $\alpha$ and added to the original Barlow Twins loss in Equation \ref{eq:bt}:

\begin{equation}
\mathcal{L_{\mathcal{BT}_{\tiny\hspace{-0.5em}\text{sparse}}}} = \mathcal{L_{\mathcal{BT}}} + \alpha \| W \|_{2,1}
\label{eq:sparse_bt}
\end{equation}

The sparsity constraint on the projector  is particularly applicable to the current setting where we consider further pre-training on smaller datasets, where the risk of overfitting will be greater which is further magnified by the large dimensionality of the projector network. Hence we use the loss in Equation \ref{eq:sparse_bt} in the curent work.


The Barlow Twins algorithm has several properties that make it potentially more applicable in the low data regime. For example, Barlow Twins achieves excellent performance when fine-tuning with limited data on ImageNet \cite{zbontar2021barlow}. As an additional point, Barlow Twins can be trained with batches as low as 128, in contrast to other self-supervised methods which may need batches above 4000  \cite{chen2020simple}. This is helpful in the case that a limited computing budget is available, e.g. using very large batches may not be possible due to memory limitations. 
 
\subsection{Framework}
\label{sec:framework}

\begin{figure*}[htbp!]
  \centering\includegraphics[scale=0.75]{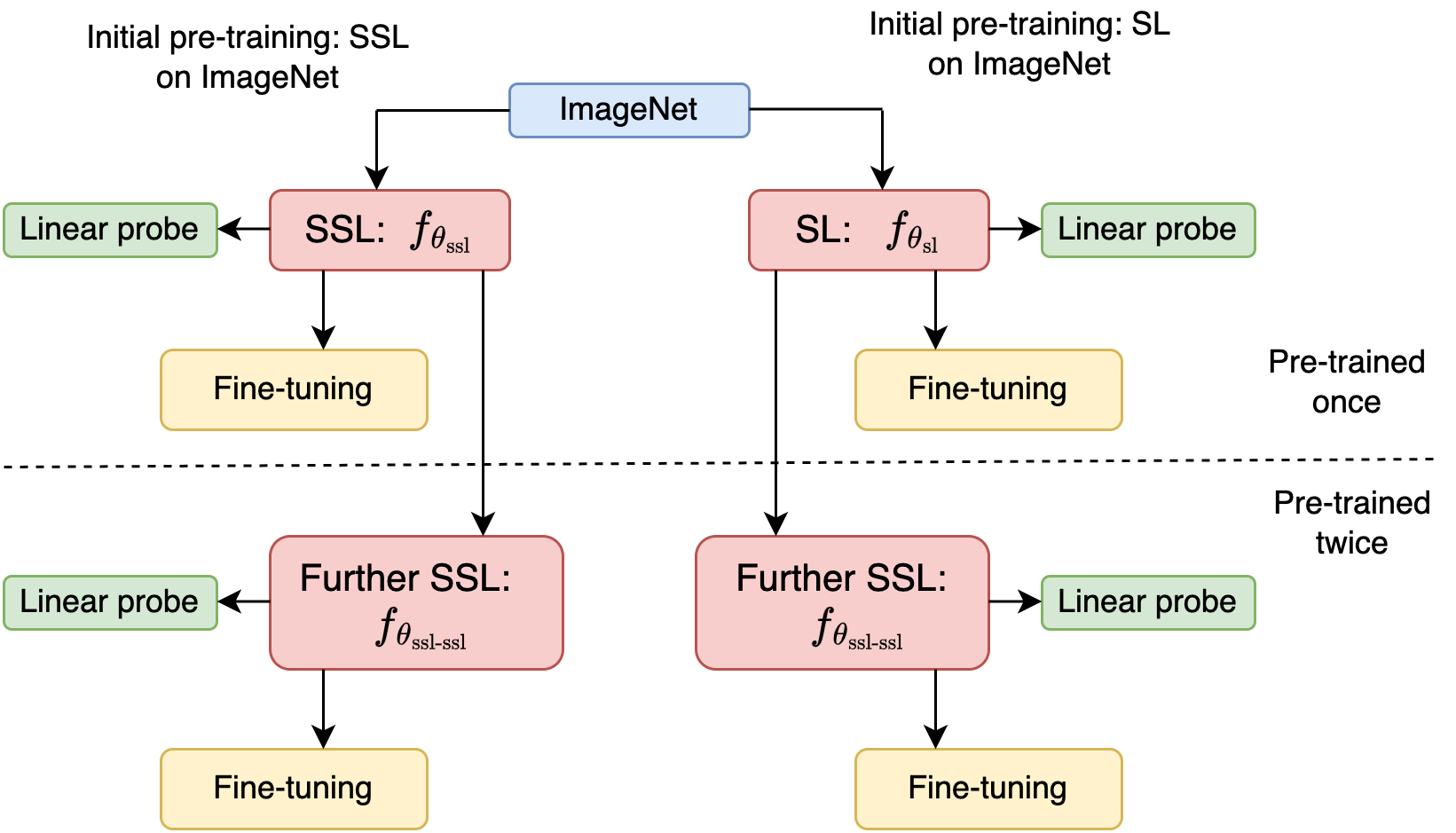}
\caption{The SSL framework shows two ResNet-50 backbone architectures pre-trained on ImageNet, either in a supervised or self-supervised fashion. We implement further self-supervised pre-training  on task-specific datasets (skin condition datasets in our case) using Barlow Twins.  We implement linear probe (in green) and full network fine-tuning (in yellow) each model weight initialisation. Note also the  figure can be cut horizontally which splits it into weights of two kinds: pre-trained once and pre-trained twice. Cutting the figure vertically splits it into results from initially supervised pre-trained (on the left) and initially self-supervised pre-trained (on the right). Note that any suitable large dataset can be utilised for the initial pre-training phase.}
\label{fig:framework}
\end{figure*}

We next provide the details of our SSL framework, outlining the major components in Figure \ref{fig:framework}. We utilise two backbone neural networks,  each having a ResNet-50 architecture and pre-trained on ImageNet, either in a supervised or self-supervised fashion. We fine-tune both backbone networks on a small labelled skin lesion dataset and compare their performance. We also train linear probes against the frozen backbones, to understand the source of fine-tuning performance differences. We then perform further self-supervised pre-training, on several different datasets.  We fine-tune and linear probe the doubly pre-trained models in the same manner as originally performed on the (pre-trained once) backbones. This enables comparison across all model weight initialisation, as shown in pink in Figure \ref{fig:framework}.

\subsubsection{ImageNet pre-trained networks}
\label{sec:ImageNet_pre-trained_networks}
ImageNet is a large annotated database of natural images \cite{ImageNet_2023} that has been prominent as a test suite for computer vision and deep learning models. It has also been popular for pre-training deep learning models \footnote{\url{https://nnabla.readthedocs.io/en/v1.39.0/python/api/models/imagenet.html}} such as VGG \cite{simonyan2014very} and ResNet   \cite{wu2019wider}.  In our framework, we use an instance of ImageNet containing approximately 1.2 million images distributed across 1000 categories \cite{russakovsky2015ImageNet,Deng2009}. Residual CNNs are deep neural networks that employ residual skip connections \cite{he2016deep}. This is an architectural design that adds the input of a neural network module to its output: $x + f(x)$, where $f$ denotes the function performed by one or more layers. This enables the training of much deeper networks, as $f(x)+x$ is easier to optimise. For instance, if an identity mapping is optimal, the network can simply learn $f(x) \approx 0$.

Our framework (Figure \ref{fig:framework}) utilises two base networks, both pre-trained on ImageNet but using different approaches. Both networks have a ResNet-50 backbone architecture \cite{he2016deep}, which we denote as the encoder $f_\theta$, where $\theta$ represents the network parameters. ResNet-50 is a residual CNN with 50 layers and an output dimensionality of 2048. We refer to the output of the encoder $f_{\theta}(x)$ as the `encoder representation', where the parameters may be frozen in which case $f_{\theta}$ acts purely as a feature extractor, or unfrozen where we can assess how well the pre-training adapts to the new task.

The two pre-training approaches are:
\begin{enumerate}
\item Supervised Learning (SL): where the pre-training architecture is: $L \circ f_\theta$, with $L$ the final linear layer \cite{torchvision_resnet50}.
\item Self-Supervised Learning (SSL): where the pre-training architecture is: $P \circ f_\theta$, where $P$ represents the projector layers, using the Barlow Twins algorithm \cite{barlowtwins}.
\end{enumerate}

More formally, let $\mathcal{D}=(X, y)$ be a dataset used for pre-training (in our case, ImageNet). We denote the pre-training processes as follows:
$f_{\theta_{\scriptscriptstyle\text{SL}}} = SL(f_\theta, \mathcal{D})$ for supervised learning and $f_{\theta_{\scriptscriptstyle\text{SSL}}} = SSL(f_\theta, X)$ for self-supervised learning. Note that SSL requires only the unlabelled data $X$, while SL uses both $X$ and $y$. Our base networks can thus be represented as: $f_{\theta_{\scriptscriptstyle\text{SL}}}$ and $f_{\theta_{\scriptscriptstyle\text{SSL}}}$, highlighting their shared architecture but different weights.

\subsection{Data for supervised learning}
Skin lesion classification is a well-studied problem using deep learning models \cite{
mahbod2019skin,zhang2019attention,lopez2017skin}. Brinker et al. \cite{brinker2019deep} demonstrated that it is possible to train models that outperform human experts, which is primarily due to the existence of large labelled datasets. Our framework is motivated by cases where such datasets do not exist, such as oral cancer. We present the data used for fine-tuning in Table \ref{tab:isic_data} where the training set only has 2,554 samples which make less than 12\% percent of total data. Note that over 90\% of oral cancer cases are squamous cell carcinomas \cite{warnakulasuriya2009global}.

\begin{table}[htbp!]
\centering
\begin{tabular}{lcc}
\toprule
\textbf{Lesion type} & \textbf{Train} & \textbf{Test}\\
\midrule
Actinic Keratosis  & 306 & 498 \\
Basal Cell Carcinoma & 500 & 2549\\
Benign Keratosis & 467 & 1663\\
Dermatofibroma     & 55 & 173\\
Melanoma           & 500 & 3339\\
Nevus              & 500 & 10601\\
Squamous Cell Carcinoma & 171 & 414\\
Vascular Lesion    & 55 & 186\\
\midrule
\textbf{Total}     &  2,554& 19,423\\
\bottomrule
\end{tabular}
\caption{Lesion types and number of samples in the training and test set, per skin cancer lesion type in the ISIC2019* dataset.}
\label{tab:isic_data}
\end{table} We obtain labelled skin lesion data from ISIC2019 \cite{andrewmvd2019isic} which is a subset of the ISIC database \cite{ISIC2023}. We call our dataset ISIC2019*, since we use only 2,554 samples for training. There are 8 lesion categories, 4 of which are benign, that are not cancerous (benign keratosis, dermatofibroma, nevus, vascular lesion) with the remainder being malignant, that is cancerous or pre-cancerous. The category actinic keratosis is precancer, whereas basal cell carcinoma, melanoma, and squamous cell carcinoma  are all cancerous. There are only 171 squamous cell carcinoma samples, making this a very challenging class imbalanced dataset.  

\subsection{Fine-tuning}
\label{sec:fine_tuning}

In our framework (Figure \ref{fig:framework}), we subsequently fine-tune the  pre-trained networks  using the training data in Table \ref{tab:isic_data}. Therefore, we implement supervised learning on $L \circ f_\theta$, where $f_\theta$ is a pre-trained encoder and $L $ is a randomly initialised linear layer mapping to the lesion categories. First, the pre-trained encoder is frozen, and the linear head is fit for 1 epoch against the frozen encoder. In this phase, we use the Adam optimiser \cite{kingma2014adam} with $lr=0.001$.  Next, we unfreeze the encoder and run a learning rate search to find a good maximum learning rate: $lr_{\text{max}}$. Lastly, we train the whole network (for 40 epochs) using the 1cycle learning rate policy and  maximum learning rate  $lr_{\text{max}}$. The 1cycle policy (Smith et al. \cite{smith2017super, smith2018disciplined}) is a learning rate scheduler, that works particularly well in low data settings. It  linearly increases the learning rate during training to a maximum value $lr_\text{max}$, and then decreases to a minimum. We provide a full discussion in the appendix \ref{sec:fine_tuning_details}. We use the Adam optimiser in all stages of training and present the full fine-tuning implementation details in Appendix \ref{sec:fine_tuning_details}.

\subsection{Linear Probe}

Linear probing (also known as linear evaluation) is a method used to assess the quality of the  representations learned by a neural network \cite{misra2019self}. In our framework (Figure \ref{fig:framework}), linear probing involves freezing the pre-trained encoder $f_\theta$ and only updating the parameters in the final linear layer $L$. In linear probing,  we use the same training procedure as during full network fine-tuning, except the encoder is frozen throughout. 

\subsection{Evaluation Metrics}

After performing supervised learning (either full network fine-tuning or linear probing), we evaluate model performance on the test dataset as described in Table \ref{tab:isic_data}. We conduct 35  independent model training and test runs using different initial random initialisation of the final linear layer $L$. We report two primary evaluation metrics, including classification accuracy and weighted-average F1 score for the test dataset.

We compute the weighted-average F1 score as follows: 
$F1_i = \sum_{j=1}^{k} w_j \cdot F1_{ij}$,
where $k$ is the number of categories, $w_j$ is the frequency of category $j$ in the test dataset, and $F1_{ij}$ is the F1 score for category $j$ in the model training run $i$.
This approach allows us to account for class imbalance in the test set while providing a robust measure of model performance across multiple runs.


\subsection{Further pre-training procedure}

\begin{figure}[htbp!]
    \centering
    \includegraphics[scale=0.85]{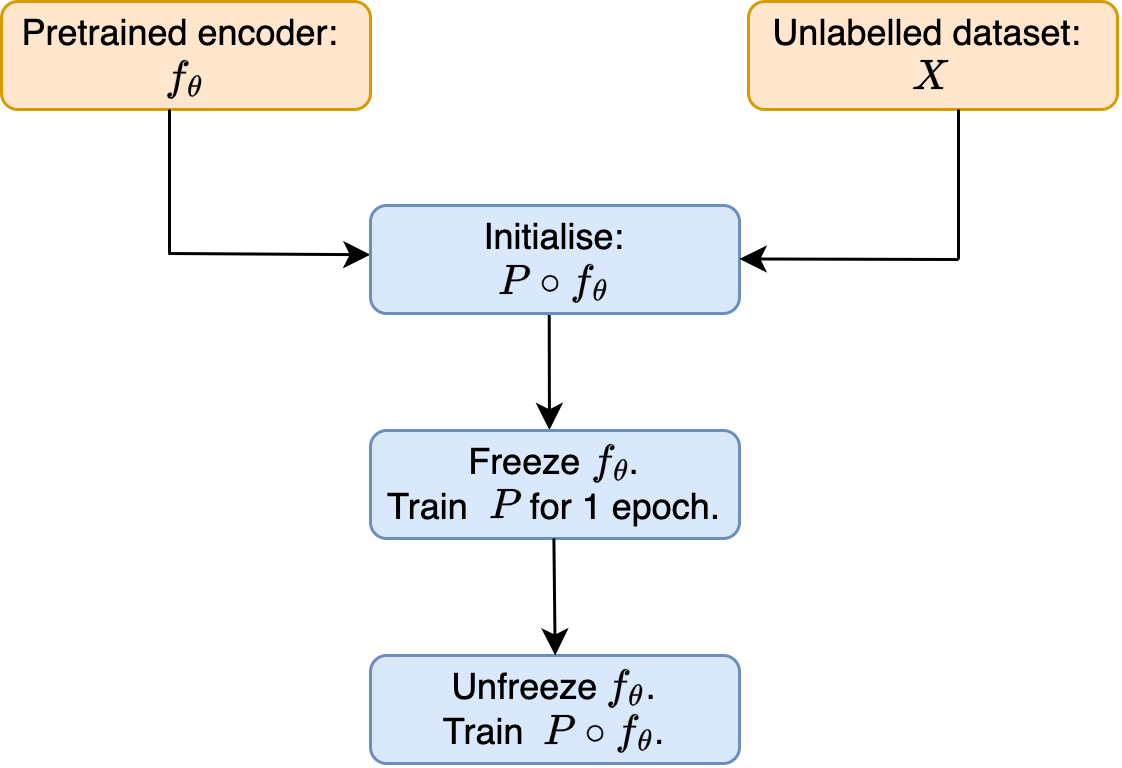}
   \caption{The procedure for further self-supervised pre-training. We reinitialise the projector and train for one epoch against the frozen encoder. We then unfreeze  the encoder and training proceeds as normal, where we train $P \circ f_\theta$  for several epochs on $X$ with SSL. We also consider unfreezing only the final part of the encoder, which we denote SSL$_p$ (partial encoder training). 
   We can view this process as a function of two inputs: a given pre-trained encoder, and an unlabelled dataset, i.e. SSL(X,$f_\theta$).}
    
    \label{fig:further_pre-training}
\end{figure}

We describe our methodology for implementing further self-supervised pre-training, which requires pre-training an (already pre-trained) encoder $f_\theta$. In Section \ref{sec:ImageNet_pre-trained_networks}, we noted that self-supervised pre-training can be viewed as a function of an unlabelled image dataset $X$, and an initialised  network with parameters  $\theta$, i.e. $f_{\theta} \leftarrow SSL(f_{\theta},X)$, where $\theta$ is randomly initialised. We specify an approach to extending the definition of $SSL(.)$ for the case when $\theta$ has already been pre-trained, which in our case will mean $\theta \in \{\theta_{\text{SL}}, \theta_{\text{SSL}} \}$. This is analogous to supervised transfer learning, whereby the training process may be different under the knowledge that the network is pre-trained, vs. being randomly initialised. 

Hence, to perform further SSL pre-training, we require a pre-trained encoder $f_\theta$ and an unlabelled image dataset $X$. Recall that when training Barlow Twins the architecture is: $P \circ f_{\theta}$ where $P$ is a projector network that is discarded after training.   Figure \ref{fig:further_pre-training} presents an overview for further pre-training, with further details as follows. 

\begin{enumerate}
\item \textbf{Projector Reinitialisation}: We initialised a new projector head $P$ comprising three layers, each with 8192 units. This fully connected network employs ReLU activation functions and batch normalisation, consistent with the original ImageNet pre-training architecture \cite{zbontar2021barlow}.

\item \textbf{Training Strategy}:
\begin{enumerate}
\item \textbf{Initial Epoch}: We freeze  the encoder $f_{\theta}$ and update  only the projector's parameters for the first epoch using the Adam optimiser, with fixed $lr=0.001$. This step prevents the random projector from immediately degrading the pre-trained features in the encoder. 
\item \textbf{Subsequent Training}: After training the projector for the first epoch, we can execute two approaches:
\begin{itemize}
\item Full encoder training: We unfreeze the entire encoder and train the entire network: $P \circ f_{\theta}$, for 100 epochs.
\item Partial encoder training: We unfreeze just the final 3 layers of the encoder while keeping the earlier layers frozen. These final 3 layers consist of a $1\times1$, $3\times3$, $1\times1$ convolution sequence, forming a bottleneck block \cite{he2016deep, torchvision_resnet50}. This approach aims to preserve low-level features while allowing the adaptation of higher-level features to the new dataset. We denote partial encoder training as SSL$_{p}$.
\end{itemize}

Both approaches employ the 1cycle policy \cite{smith2017super,smith2018disciplined} as the learning rate schedule, consistent with the fine-tuning process described in sections \ref{sec:fine_tuning} and \ref{sec:fine_tuning_details}. This policy is particularly effective in limited data scenarios, making it well-suited for our SSL framework to leverage relatively small datasets; SSL typically involves very large datasets \cite{lecun2021self}.
\end{enumerate}
\end{enumerate}

Hence, we have an approach that extends the definition of $SSL(.)$ as in $f_{\theta} \leftarrow SSL(f_{\theta},X)$ to work when $\theta$ is pre-trained rather than randomly initialised. The approach to further pre-training is inspired by supervised transfer learning techniques that aim to leverage the pre-trained ResNet-50 ImageNet features while adapting the model to the new dataset. Essentially, we use a discriminative learning rate scheme, whereby the encoder has a learning rate of zero for the first epoch and a non-zero learning rate thereafter. For partial encoder training SSL$_p$, the learning rate remains zero throughout except for the final bottleneck block.

By comparing full and partial encoder fine-tuning, we can assess the trade-off between preserving pre-trained features and adapting to the new data distribution.  
We set the hyperparameter $\lambda=\frac{1}{8192}$ and the regularisation hyperparameter $\alpha=0.01$ (Equation \ref{eq:sparse_bt}). We adopt the data augmentation strategy from previous papers \cite{zbontar2021barlow,grill2020bootstrap}, and present a representative sample in Figure \ref{fig:isic_bt_cancer}. The full implementation details are in  Appendix \ref{sec:bt_implementation} and on GitHub \footnote{\url{https://github.com/hamish-haggerty/base-rbt}}


\subsection{Data for further self-supervised pre-training}
\label{sec:further_pre-training}

We utilise 3 base datasets for further self-supervised pre-training:  DermNet \cite{dermnet,dermnet_kaggle}, UFES20 \cite{pacheco2020pad}, and a subset of ISIC2019 data itself \cite{ISIC2023,andrewmvd2019isic} as summarised in Table \ref{tab:additional_ssl_data}. The base datasets feature different types of skin conditions and provide  a similar data distribution to the fine-tuning data in Table \ref{tab:isic_data}.

We combined these base datasets   to form 3 datasets for further pre-training: I, IU, IUD, as summarised in Table \ref{tab:combined_additional_ssl_data}. We note that I refers to the fine-tuning data as shown in Table \ref{tab:isic_data} and IU refers to all the base datasets combined -  except for DermNet. IUD is IU plus DermNet (i.e., all the base datasets together). 



\begin{table*}[htbp!]
\centering
\begin{tabular}{@{}p{2cm}p{2cm}p{8cm}@{}}
\toprule
\textbf{Dataset} & \textbf{Image Count} & \textbf{Description} \\ \midrule
\addlinespace[0.5em]
DermNet & 19,559 & Covers a wide range of general skin diseases with 23 categories, some overlapping with ISIC categories such as melanoma. \\
\addlinespace[0.5em]
UFES20 & 2,298 & Similar to ISIC, includes skin lesion images with six categories: Basal Cell Carcinoma, Squamous Cell Carcinoma, Actinic Keratosis, Seborrheic Keratosis, Melanoma, and Nevus. \\
\addlinespace[0.5em]
ISIC$_{\text{fine-tune}}$ & 2,554 & Data used for fine-tuning as in Table \ref{tab:isic_data}. \\
\addlinespace[0.5em]
ISIC$_{\text{val}}$ & 1,280 & Additional ISIC2019 samples  \\ 
\addlinespace[0.5em]
\bottomrule
\end{tabular}
\caption{Summary of base datasets involved in further pre-training. We combine these base datasets to build 3 datasets for further pre-training, as summarised in Table \ref{tab:combined_additional_ssl_data}.  Note that in the case of further SSL pre-training, none of the labels in the above datasets are used.}
\label{tab:additional_ssl_data}
\end{table*}

\begin{table}[htbp!]
\centering
\begin{tabular}{@{}lcc@{}}
\toprule
\textbf{Dataset} & \textbf{Image Count} & \textbf{Description} \\ \midrule
I                & 2,554 &   ISIC$_{\text{fine-tune}}$  
\\
IU               & 6,132       & I+ISIC$_{\text{val}}$+ UFES20 \\ 
IUD              & 31,823      & IU + DermNet   \\ \bottomrule
\end{tabular}
\caption{The three datasets used for further pre-training based on the  combinations of the base datasets in Table \ref{tab:additional_ssl_data}. The I dataset refers to  the ISIC fine-tuning data: i.e. I=ISIC$_{\text{fine-tune}}$. The IU dataset is I combined with selected additional ISIC validation data (ISIC$_{\text{val}}$) and the UFES20 dataset. Finally, IUD is the IU dataset combined with  DermNet. We oversample IU at ratio of 2:1, hence IUD contains two copies of IU and one copy of DermNet.}
\label{tab:combined_additional_ssl_data}
\end{table}

\section{Results}
\label{sec:results}


We use the skin lesion dataset in Table \ref{tab:isic_data} for experiments (whether full network fine-tuning or linear probing). We use a ResNet-50 encoder $f_{\theta}$ as the backbone architecture in all the experiments, with variations in how the weights $\theta$ have been pre-trained. In the final section we present results for an oral cancer histopathology dataset.

\subsection{Pretrained once on ImageNet: supervised vs. self-supervised pre-training}
\label{sec:isic_result_base}


\begin{table*}[htbp!]
\centering
\begin{tabular}{lllcccc}
\toprule
\textbf{Initial} & \textbf{Further} & \textbf{Further} & \textbf{Accuracy} & \textbf{Accuracy} & \textbf{F1} & \textbf{F1} \\
\textbf{Pre-training} & \textbf{Pre-training} & \textbf{Pre-training} & \textbf{(mean)} & \textbf{(std)} & \textbf{(mean)} & \textbf{(std)} \\
& \textbf{Dataset} & \textbf{Method} & & & & \\ 
\midrule
\multicolumn{7}{c}{\textit{Pre-trained Once on ImageNet}} \\
\midrule
SL & - & - & 0.65701 & 0.02109 & 0.66873 & 0.01710 \\
SSL & - & - & 0.69112 & 0.01117 & 0.69628 & 0.00902 \\
\midrule
\multicolumn{7}{c}{\textit{Pre-trained Twice}} \\
\midrule
\multirow{4}{*}{SL} & IUD & SSL & 0.69565 & 0.00954 & 0.70213 & 0.00814 \\
 & IUD & SSL$_p$ & 0.69809 & 0.01413 & 0.70484 & 0.01213 \\
 & IU & SSL$_p$ & 0.69686 & 0.01093 & 0.70373 & 0.00919 \\
 & I & SSL$_p$ & 0.70065 & 0.01246 & 0.70619 & 0.01038 \\
\midrule
\multirow{4}{*}{SSL} & IUD & SSL & 0.69852 & 0.00893 & 0.70550 & 0.00746 \\
 & IUD & SSL$_p$ & 0.69734 & 0.01002 & 0.70499 & 0.00837 \\
 & IU & SSL$_p$ & 0.69778 & 0.00923 & 0.70406 & 0.00784 \\
 & I & SSL$_p$ & 0.69583 & 0.00925 & 0.70297 & 0.00795 \\
\bottomrule
\end{tabular}
\caption{
Fine-tuning results for models pre-trained once and twice based on the   ISIC2019* dataset presented in Table \ref{tab:isic_data}.
SSL$_p$ indicates partial encoder training during further pre-training (which is the same as SSL except the encoder is frozen aside from the final 3 layers).  I, IU, and IUD refer to different datasets used for further pre-training as defined in Table \ref{tab:combined_additional_ssl_data}. We report the mean and standard deviation (std) of 35 independent model training runs showing classification accuracy and weighted-average F1 score.}
\label{tab:combined_isic_semi_supervised_results}
\end{table*}

\begin{table}[htbp!]
\centering
\begin{tabular}{lcccc}
\toprule
\textbf{Training} & \textbf{Accuracy} & \textbf{Accuracy} & \textbf{F1} & \textbf{F1} \\
\textbf{time} & \textbf{(mean)} & \textbf{(std)} & \textbf{(mean)} & \textbf{(std)} \\ 
\midrule
 40 epochs  & 0.35802 & 0.01834 & 0.38614 & 0.01665 \\
 80 epochs  & 0.36460 & 0.02142 & 0.38819 & 0.02379 \\
 160 epochs & 0.38537 & 0.01925 & 0.41055 & 0.02004 \\
\bottomrule
\end{tabular}
\caption{Results for the ISIC2019 dataset  using random weight initialisation for  ResNet-50 models. The setup is similar to Table \ref{tab:combined_isic_semi_supervised_results} except we use random weight initialisation, instead of pre-trained weights. We provide results (accuracy and F1) of  5 model training runs only.}
\label{tab:random_weights_results}
\end{table}

We begin by comparing the fine-tuning results of the backbone ResNet-50 encoders that have been pre-trained on ImageNet as defined in Section \ref{sec:ImageNet_pre-trained_networks}. We recall that these are the base models that have been pre-trained once on ImageNet with either SL or SSL. Hence, the models to be fine-tuned have the form: $L \circ f_\theta$, where $\theta \in \{\theta_{\text{SL}}, \theta_{\text{SSL}} \}$ and $L$ is a randomly initialised linear layer. We use the  dataset in Table \ref{tab:isic_data} for fine-tuning  and present the results  in Table \ref{tab:combined_isic_semi_supervised_results}. Note that the randomness  (weight initialisation) across the independent model training runs comes entirely from $L$, i.e. the linear layer. 
\enlargethispage{-\baselineskip}
The main observation is that SSL pre-training outperforms SL pre-training in terms of both accuracy and average weighted F1 score. Table \ref{tab:isic_semi_sup_sig} demonstrates that this difference is highly statistically significant. In other words, pre-training on ImageNet using Barlow Twins yields better fine-tuning performance compared to supervised pre-training on ImageNet. We can denote this result as $f_{\theta_{\scriptscriptstyle\text{SSL}}}\underset{\scriptscriptstyle\text{(fine-tune)}}{>} f_{\theta_{\scriptscriptstyle\text{SL}}}$, where the subscript `fine-tune' indicates that the comparison is based on fine-tuning performance.\footnote{Formally, $\underset{\scriptscriptstyle\text{(fine-tune)}}{>}$ defines a partial order on models. For a set of pre-trained models, we fine-tune each model and record its mean test accuracy (say). Then, $f_{\theta}\underset{\scriptscriptstyle\text{(fine-tune)}}{>} f_{\theta'}$ holds if and only if the mean test accuracy of $f_{\theta}$ (i.e. after fine-tuning the model with initial weights $\theta$) is significantly higher than that of $f_{\theta'}$, also post fine-tuning.} 
This result raises a question: why does self-supervised pre-training lead to superior performance? We explore two possible hypotheses in the following section.

\subsection{Linear probe analysis of base models}

\begin{table*}[htbp!]
\centering
\begin{tabular}{lllcccc}
\toprule
\textbf{Initial} & \textbf{Further} & \textbf{Further} & \textbf{Accuracy} & \textbf{Accuracy} & \textbf{F1} & \textbf{F1} \\
\textbf{Pre-training} & \textbf{Pre-training} & \textbf{Pre-training} & \textbf{(mean)} & \textbf{(std)} & \textbf{(mean)} & \textbf{(std)} \\
& \textbf{Dataset} & \textbf{Method} & & & & \\ 
\midrule
\multicolumn{7}{c}{\textit{Pre-trained Once on ImageNet}} \\
\midrule
SL & - & - & 0.59281 & 0.01097 & 0.61513 & 0.00817 \\
SSL & - & - & 0.66278 & 0.00686 & 0.66752 & 0.00503 \\
\midrule
\multicolumn{7}{c}{\textit{Pre-trained Twice}} \\
\midrule
\multirow{3}{*}{SL} & IUD & SSL$_p$ & 0.67033 & 0.00667 & 0.67838 & 0.00533 \\
 & IU & SSL$_p$ & 0.68370 & 0.00675 & 0.68946 & 0.00616 \\
 & I & SSL$_p$ & 0.67588 & 0.00582 & 0.68170 & 0.00476 \\
\midrule
\multirow{3}{*}{SSL} & IUD & SSL$_p$ & 0.68364 & 0.00430 & 0.68826 & 0.00363 \\
 & IU & SSL$_p$ & 0.68645 & 0.00430 & 0.69083 & 0.00329 \\
 & I & SSL$_p$ & 0.69403 & 0.00404 & 0.69671 & 0.00365 \\
\bottomrule
\end{tabular}
\caption{
Linear probe results for  ResNet-50 models  pre-trained once and twice using  ISIC2019* dataset. 
SSL$_p$ indicates partial encoder training during further pre-training. I, IU, and IUD refer to different datasets used for further pre-training as defined in Table \ref{tab:combined_additional_ssl_data}. We report the mean and standard deviation (std) of 35 independent model training runs, showing classification accuracy and weighted-average F1 score.}
\label{tab:combined_isic_linear_probe_results}
\end{table*}

\begin{table}[htbp!]
\centering
\begin{tabular}{lcc}
\toprule
\textbf{Pre-training} & \textbf{Linear probe} & \textbf{Fine-tuning} \\
\textbf{Method} & \textbf{(Mean Accuracy)} & \textbf{(Mean Accuracy)} \\ 
\midrule
SL & 0.59281 & 0.65701 \\
SSL & 0.66278 & 0.69112 \\
\bottomrule
\end{tabular}
\caption{Comparison of accuracy for linear probe and fine-tuning for ResNet-50 models \textbf{pre-trained once} on ImageNet. The results for the F1 score show similar trends.}
\label{tab:comparison_table}
\end{table}

We employ linear probing for assessing representation quality by training only the final linear layer on frozen features, in order to evaluate the fine-tuning performance difference between SL and SSL pre-trained networks. Hence, the experimental setup is exactly the same as  Section \ref{sec:isic_result_base}, except that we freeze the backbone encoders $f_\theta$ and only update the parameters of the linear layer $L$  during training. We consider two hypotheses to explain the superior fine-tuning performance of SSL:

 \begin{enumerate}
 
  \item The  \textit{fast adaptation hypothesis} posits that there is no significant difference in linear probe performance between SSL and SL. This would suggest that the SSL weights are more amenable to rapid learning on new data during fine-tuning.
 
  \item \textit{The feature reuse hypothesis} proposes that SSL outperforms SL in linear probe performance. This would indicate that the self-supervised pre-training produces more reusable features, explaining the superior fine-tuning performance. We adapted the terminology `rapid learning' and `feature reuse' from Raghu et al. \cite{raghu2020rapid}.
    
 \end{enumerate}

The results in Table \ref{tab:combined_isic_linear_probe_results} strongly support the feature reuse hypothesis, where the frozen SSL representations significantly outperform their SL counterparts, with accuracies of 66.3\% and 59.3\%, respectively. The \textit{linear probe} performance of SSL is comparable to the \textit{fine-tuning} performance of SL. To make this clear we include a Table \ref{tab:comparison_table} with a side-by-side comparison of linear probe and fine-tuning performance. Hence,  we can summarise this result as $f_{\theta_{\scriptscriptstyle\text{SSL}}}\underset{\scriptscriptstyle\text{(linear probe)}}{>} f_{\theta_{\scriptscriptstyle\text{SL}}}$. 


In order to better understand the (pre-trained once) results in Table \ref{tab:combined_isic_linear_probe_results}  we recall how each base  ResNet-50 encoder is pre-trained. The SL model is trained on a supervised classification task with 1000 categories from ImageNet. This pre-training process can be represented as $L_{1000} \circ f_\theta$, where $f_\theta$ is the encoder (feature extractor) and $L_{1000}$ is a linear layer that maps the encoder's output to 1000 class scores. Specifically, for an input image $x$, the encoder produces a feature vector $r = f_\theta(x)$. The linear layer $L_{1000}$ then computes scores for each of the 1000 ImageNet categories as linear combinations of this feature vector. We can represent this computation as $s = L_{1000}r$, where $s$ is the vector of scores for all categories. Suppose the $i$-th category corresponds to "dog", and let the $i$-th row of $L_{1000}$ be $w_i^T$, then the score for "dog" would be computed as $s_i = w_i^T r$. After pre-training, $L_{1000}$ is discarded, leaving us with the pre-trained encoder: $f_{\theta_{\scriptscriptstyle\text{SL}}}$. When we perform linear probing on our new task (skin lesion classification), we have $L \circ f_{\theta_{\scriptscriptstyle\text{SL}}}$, where $L$ is a new linear layer that maps to our 8 skin lesion categories, and the parameters in $\theta_{\scriptscriptstyle\text{SL}}$ are frozen. 

The difficulty arises because the features learned by $\theta_{\scriptscriptstyle\text{SL}}$ are optimised to discriminate between the 1000 ImageNet categories. These features are trained such that linear combinations of them can effectively distinguish between categories such as  "dog", "cat", "car", etc. However, such features may not be as effective for distinguishing between different types of skin lesions. It is perhaps not surprising that linear combinations of such features may not easily translate to our  task of skin lesion classification. In contrast, Barlow Twins SSL pre-training does not use any class-specific information. Instead, it learns to create similar representations for different augmented views of the same image. This approach likely produces more abstract and generalisable features that are not tied to specific ImageNet categories. This fundamental difference in the nature of learned  features appears to be the primary reason for the superior performance of $f_{\theta_{\scriptscriptstyle\text{SSL}}}$, in both linear probe and fine-tuning. Our results suggest that SSL pre-training on ImageNet via Barlow Twins yields a more versatile and transferable feature representation when compared to supervised pre-training, particularly when adapting to new tasks like skin lesion classification.

These considerations motivate the question of whether it is possible to perform a minimal amount of further self-supervised pre-training to enhance the SL encoder representation. In the following section, we show that minimal further pre-training of $f_{\theta_{\scriptscriptstyle\text{SL}}}$ boosts both linear probe and fine-tuning performance.

\subsection{Fine-tuning results after further pre-training}

We present the fine-tuning results following further pre-training in Table \ref{tab:combined_isic_semi_supervised_results}. Further pre-training on the IUD dataset boosts performance for both (initially) SL and SSL pre-trained models. This improvement is particularly noticeable for SL, where we observe a substantial performance gain in terms of accuracy and F1 scores. In the case of SSL, although the improvement is small, it remains statistically significant, as shown in Table \ref{tab:isic_semi_sup_sig}.
We found no significant difference in performance between pre-training the entire network on IUD (SSL) and pre-training only the final three layers (SSL$_p$). Hence, for subsequent pre-training experiments, we only considered the training of the final bottleneck block. 

Another noteworthy observation is the comparable performance achieved when pre-training on datasets of vastly different sizes. Specifically, pre-training on the IUD dataset (containing 25,691 images), the IU dataset, and the I dataset alone (with only 2,554 images) yielded similar results. This finding suggests that the relevance of the pre-training data may be more crucial than its quantity. Perhaps most surprisingly, after further self-supervised pre-training of each of the base models, we observed no subsequent performance difference. In order to understand this better, we recall the initial result that SSL pre-training outperformed SL pre-training, i.e. $f_{\theta_{\scriptscriptstyle\text{SSL}}}\underset{\scriptscriptstyle\text{(fine-tune)}}{>} f_{\theta_{\scriptscriptstyle\text{SL}}}$. However, if we further pre-train each base model (i.e. $f_{\theta_{\scriptscriptstyle\text{SL}}}$ and $f_{\theta_{\scriptscriptstyle\text{SSL}}}$)  this inequality becomes an equality: $f_{\theta_{\scriptscriptstyle\text{SSL-SSL}}}\underset{\scriptscriptstyle\text{(fine-tune)}}{\approx} f_{\theta_{\scriptscriptstyle\text{SL-SSL}}}$. This holds even when using the fine-tuning data for further pre-training, which in our case is just 2,554 samples.


\subsection{Linear probe analysis after further pre-training}

\begin{table}[htbp!]
\centering
\begin{tabular}{lcc}
\toprule
\textbf{Initial} & \textbf{Linear probe} & \textbf{Fine-tuning} \\
\textbf{Pre-training} & \textbf{(Mean Accuracy)} & \textbf{(Mean Accuracy)} \\ 
\midrule
SL & 0.67588 & 0.70065 \\
SSL & 0.69403 & 0.69583 \\
\bottomrule
\end{tabular}
\caption{Comparison of accuracy for linear probe and fine-tuning on ISIC2019* after further pre-training. SL denotes supervised learning, and SSL denotes self-supervised learning (Barlow Twins) for initial pre-training on ImageNet. Both models implemented identical further pre-training on the fine-tuning dataset (I=ISIC2019*) using SSL$_p$ (partial encoder training).
}
\label{tab:comparison_table_after_pretraining}
\end{table}

Our earlier linear probe experiments demonstrated that feature reuse largely explained the fine-tuning performance difference between SL and SSL pre-training. To investigate this phenomenon further, we conducted similar linear probe experiments on the frozen representations after further pre-training. The results in Table \ref{tab:combined_isic_linear_probe_results} reveal  a significant increase in linear probe performance following further pre-training of the SL encoder. Surprisingly, after further pre-training of $f_{\theta_{\scriptscriptstyle\text{SL}}}$, the representation outperformed the $f_{\theta_{\scriptscriptstyle\text{SSL}}}$ representation. In general, for any choice of dataset for further pre-training,   we found   that: 
$f_{\theta_{\scriptscriptstyle\text{SL-SSL}}}\underset{\scriptscriptstyle\text{(linear probe)}}{>} f_{\theta_{\scriptscriptstyle\text{SSL}}}$.
We recall our prior result that: $f_{\theta_{\scriptscriptstyle\text{SSL}}}\underset{\scriptscriptstyle\text{(linear probe)}}{>} f_{\theta_{\scriptscriptstyle\text{SL}}}$. Despite this initial gap in encoder representation, it appears that even minimal further pre-training of the SL encoder on the fine-tuning data is sufficient to boost its representation capacity to that of the SSL encoder. Therefore, although there is a gap in the feature extraction capacity of the base encoders (as measured by linear probe performance), further pre-training the worse base model  (SL) just on the fine-tuning data is enough for this gap to close, and even reverse. 

Further pre-training of the SSL encoder also improved its linear probe performance. For instance, further pre-training of just the final 3 layers on the fine-tuning data increased linear probe accuracy from 66.3\% to 69.4\%, with a similar improvement in average weighted F1 score. After further pre-training the SSL encoder, we observed no significant difference between linear probe and fine-tuning performance (summarised in Table \ref{tab:comparison_table_after_pretraining}). This indicates that further self-supervised pre-training enhances the encoder's representation to such an extent that a simple linear classifier performs comparably to full network fine-tuning. These findings together provide strong evidence that the general trend of increased performance after further pre-training is primarily due to enhanced feature reuse in the encoder representation. 

\subsection{Validation on oral cancer dataset}


We finally consider an oral cancer histopathology dataset, which can be seen in Table \ref{tab:oral_cancer_data} \cite{fasilkebede2023oral}. This is a relatively small binary classification dataset, consisting of only 122 labelled samples for fine-tuning: 60 normal samples and 62 oral squamous cell carcinoma (OSCC). As earlier, we first compare the fine-tuning performance of the SL and SSL pre-trained encoders. We observe the same pattern of results as for ISIC2019* data, where SSL outperforms SL which can be seen in Table \ref{tab:semi_sup_oral}. Next we consider further pre-training of the SL encoder. In the case of ISIC2019*, we could reach maximum relative performance by just pre-training on the available fine-tuning data, which was 2,554 samples. The fine-tuning data available for oral cancer histopathology classification is only 122 samples, which is very low for self-supervised learning. We increase the dataset for pre-training by adding an additional 2374 `normal' samples, as shown in Table \ref{tab:oral_cancer_data}. Note that the collection of additional normal samples compared to OSCC is significantly easier (i.e. no biopsy diagnosis is needed). We oversample the OSCC samples at a 10:1 ratio when pre-training.

Despite only having 62 OSCC samples available for pre-training, the  results are similar to the ISIC2019* data, with $f_{\theta_{\scriptscriptstyle\text{SL-SSL}}}$ having a similar performance when compared to $f_{\theta_{\scriptscriptstyle\text{SSL}}}$, and substantially higher than $f_{\theta_{\scriptscriptstyle\text{SL}}}$. In other words, while there is a substantial (fine-tuning) performance gap between the SL and SSL encoders, after further pre-training the SL encoder, this gap becomes much smaller. It is surprising this is possible even with an extremely low number (62) of malignant samples.

\begin{table}[H]
\centering
\begin{tabular}{@{}lcc@{}}
\toprule
\textbf{Dataset} & \textbf{Category} & \textbf{Image Count} \\ \midrule
\multirow{2}{*}{Fine-tuning} & Normal & 60 \\
                             & OSCC   & 62 \\ \midrule
\multirow{2}{*}{Pre-training} & Normal   & 60+2374 \\
                             & OSCC   & 62 $\times$ 10 \\ \midrule
\multirow{2}{*}{Testing} & Normal & 59 \\
                         & OSCC   & 187 \\ \bottomrule
\end{tabular}
\caption{Oral cancer histopathology dataset where we have   60 (Normal) and 62 (OSCC) samples for fine-tuning. In pre-training, we use the fine-tuning data plus some additional 'normal' samples. We oversample (10:1) the OSCC data   during pre-training.}
\label{tab:oral_cancer_data}
\end{table}



\begin{table*}[htbp!]
\centering
\begin{tabular}{lllcccc}
\toprule
\textbf{Initial} & \textbf{Further} & \textbf{Further} & \textbf{Accuracy} & \textbf{Accuracy} & \textbf{F1} & \textbf{F1} \\
\textbf{Pre-training} & \textbf{Pre-training} & \textbf{Pre-training} & \textbf{(mean)} & \textbf{(std)} & \textbf{(mean)} & \textbf{(std)} \\
& \textbf{Dataset} & \textbf{Method} & & & & \\ 
\midrule
\multicolumn{7}{c}{\textit{Pre-trained Once}} \\
\midrule
SL & - & - & 0.66574 & 0.03667 & 0.68880 & 0.03375 \\
SSL & - & - & 0.74762 & 0.02414 & 0.74204 & 0.01473 \\
\midrule
\multicolumn{7}{c}{\textit{Pre-trained Twice}} \\
\midrule
SL & Oral Cancer & SSL$_p$ & 0.73577 & 0.03497 & 0.74085 & 0.02602 \\
\bottomrule
\end{tabular}
\caption{Fine-tuning results for oral cancer histopathology where the data for fine-tuning and further pre-training is outlined in Table \ref{tab:oral_cancer_data}. SSL$_p$ indicates partial encoder training during further pre-training. We report the mean and standard deviation (std) of 35 independent model training runs for classification accuracy and weighted-average F1 scores.}
\label{tab:semi_sup_oral}
\end{table*}

\begin{figure}[h!]
    \centering
    \includegraphics[scale=0.4]{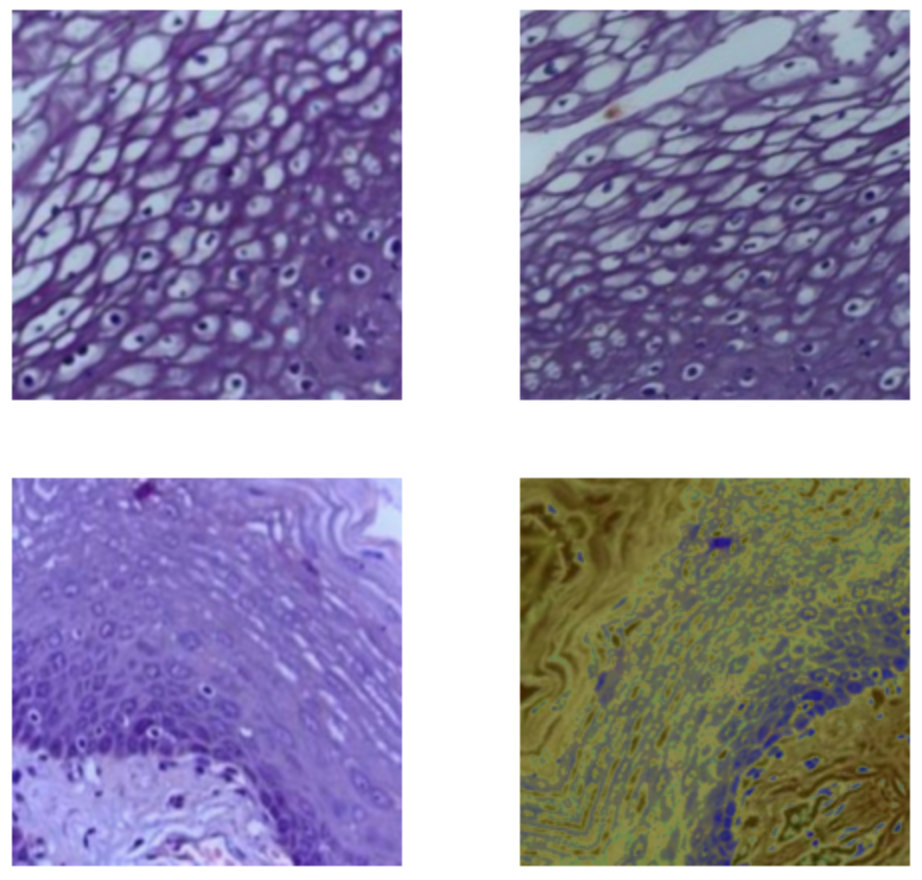}
    \caption{SSL Barlow Twins augmentations on oral cancer histopathology data where we train the network to ignore the row-wise image distortions in a non-redundant way.}
    \label{fig:bt_cancer}
\end{figure}

\section{Discussion}

In our study, we compared two ResNet-50 base networks that had been pre-trained on ImageNet using  SL and  SSL (via Barlow Twins). We fine-tuned both networks on a small skin lesion dataset and found that SSL pre-training led to superior performance.  Then we performed linear probe experiments, training just the final linear layer, with the SSL network similarly superior.  These results suggest that SSL pre-training on natural image datasets such as ImageNet produces more transferable features out of the box, while the final few layers of the SL pre-trained network may overfit to the pre-training task of classification of 1000 ImageNet categories. These considerations motivated the strategy of further SSL pre-training, particularly to enhance the SL pre-trained model.

We further pre-trained both base networks on several task-specific datasets of different sizes, including only the available fine-tuning data. We found that doing so generally enhanced performance, and using the fine-tuning data (and further pre-training the final 3 layers of the network) was equivalent to using larger datasets. Significantly, after further pre-training, the performance gap between the base networks disappeared. The SL encoder was significantly enhanced and the SSL encoder was mildly enhanced, after being further pre-trained. This was somewhat surprising as there was a substantial initial performance gap, resulting from large-scale pre-training on ImageNet. Although SL pre-training overfits the final few layers to the pre-training task, we found that minimal further pre-training is enough to undo this. 

One can view further pre-training as a kind of `pre-processing step' before fine-tuning, except the pre-processing is on the model weights, instead of the data itself. More formally, given a training dataset for fine-tuning $\mathcal{D} = (X,y)$, it is common to first normalise the features in $X$ e.g. to have mean 0 and standard deviation of 1; note that this is a function just of $X$, i.e. $X \leftarrow \text{norm}(X)$.
Similarly, given a pre-trained network $f_\theta$ (where $L \circ f_\theta$ is to be fine-tuned on $\mathcal{D}$), one can instead perform a `pre-processing' step by further pre-training $f_\theta$ on $X$. This is now a function of both the model weights and the unlabelled data $X$, i.e. $f_\theta \leftarrow SSL(f_\theta,X)$ but can still be viewed as a pre-processing step since the labels $y$ are not used. We note this and recall the initial results comparing the base models: $f_{\theta_{\scriptscriptstyle\text{SSL}}}\underset{\scriptscriptstyle\text{(fine-tune)}}{>} f_{\theta_{\scriptscriptstyle\text{SL}}}$, where SSL pre-training outperformed SL pre-training. However, after further pre-training both base models on $X$, we have: $f_{\theta_{\scriptscriptstyle\text{SSL-SSL}}}\underset{\scriptscriptstyle\text{(fine-tune)}}{\approx} f_{\theta_{\scriptscriptstyle\text{SL-SSL}}}$. Hence, the initial performance gap between SL and SSL pre-trained models closes after targeted, further pre-training, even when $X$ is small. We performed additional linear probe experiments to understand this result. After further pre-training of both base networks, the linear probe performance was significantly higher for each. In fact, after further pre-training the SSL encoder, linear probe and full network fine-tuning gave equivalent results. This was somewhat surprising since the network had already been pre-trained with SSL via Barlow Twins using the entire ImageNet dataset, which features about 1.2 million samples. Apparently, further pre-training only the final 3 layers using only  2,554 samples is enough to enhance the encoder representation such that linear probe and full network fine-tuning give similar results.

Our findings present a simple method for enhancing supervised pre-trained models. Since supervised pre-training on ImageNet is a ubiquitous technique in medical image classification \ref{sec:cancer_lit}, the findings are broadly applicable. The results also challenge the necessity of large-scale self-supervised pre-training on datasets like ImageNet - at least in the setting, where there is limited medical image data available for downstream fine-tuning. This approach opens up new possibilities for efficient model development in resource-constrained environments, which is especially relevant in medical imaging where large labelled datasets are often unavailable or inaccessible due to privacy concerns, high collection costs, or the rarity of certain conditions. Our approach also offers a streamlined method for evaluating new SSL techniques. Instead of extensive pre-training on ImageNet, researchers can assess a new SSL method's effectiveness by performing minimal further pre-training on a small dataset, using ImageNet SL weights as initialisation. This approach could substantially reduce the computational resources and time needed to evaluate new SSL techniques, potentially accelerating innovation in SSL pre-training for downstream medical image classification.

A limitation of this work is it only involved the ResNet-50 architecture, due to the availability of ImageNet pre-trained models. Future work could investigate SSL pre-training for other architectures, such as vision transformers \cite{dosovitskiy2020image}. This is of particular interest since recent work has found that on some medical classification problems, vision transformers can outperform CNNs when both are pre-trained on ImageNet \cite{diagnostics13020178}. Moreover, vision transformers appear to benefit particularly from scale. Dosovitskiy et al. \cite{dosovitskiy2020image} found that vision transformers can match or exceed ResNet transfer performance when pre-trained on datasets larger than ImageNet. Investigating how our findings extend to these more recent architectures, could provide further insights into efficient pre-training strategies for medical image analysis.

Given the proven benefit of model scale in many contexts \cite{kaplan2020scalinglawsneurallanguage,dosovitskiy2020image}, it's natural to wonder if using a larger ResNet architecture for ImageNet SSL pre-training leads to performance gains. Unfortunately, at present there are no open-source (ImageNet pre-trained) Barlow Twins models larger than a ResNet-50. However, there exist larger models pre-trained with VICReg \cite{vicreg}. VICReg is a self-supervised learning algorithm that is very similar to Barlow Twins and performs comparably on ImageNet transfer \cite{vicreg}. We fine-tuned a ResNet-50 on  ISIC2019* dataset that had been pre-trained with VICReg and found it performed identically to Barlow Twins, as expected. To explore the effect of increasing model size, we then considered fine-tuning a VICReg pre-trained ResNet-200 (x2) \cite{vicreg}. This network has an encoder dimension of 4096 and about 250 million parameters, compared to 23 million parameters for a ResNet-50, making it an order of magnitude larger in parameter count. Interestingly, the performance was the same as for the ResNet-50. This suggests that the performance of such SSL pre-training methods may be saturated with respect to the ResNet architecture and ImageNet - at least in limited data scenarios. Future work could explore whether this saturation holds for other architectures or when conducting pre-training on datasets other than ImageNet.

\section{Conclusions}
We compared two ResNet-50 base networks that had been pre-trained on ImageNet using supervised learning and self-supervised learning on a small labelled skin lesion dataset. We found that self-supervised pre-training led to better fine-tuning and linear probe performance. We demonstrated that further self-supervised pre-training on small, task-specific datasets can enhance initially supervised pre-trained models to match the performance of those initially pre-trained via self-supervised methods on large datasets such as ImageNet.  This approach could significantly reduce
computational requirements for developing effective models in limited-data scenarios.

Our study contributes to the growing body of research on transfer learning and self-supervised learning, applied
to medical image classification. As the field of artificial intelligence for medical applications continues to evolve, techniques that can maximize performance with limited data will be crucial in expanding the reach and impact of these technologies, particularly in resource-constrained healthcare settings.
 
\subsection*{Code and Data}

Our code is open source on GitHub \footnote{\url{https://github.com/hamish-haggerty/base_rbt}.} 

 

\bibliographystyle{cas-model2-names}

\bibliography{cas-refs}

\begin{thebibliography}{80}
\expandafter\ifx\csname natexlab\endcsname\relax\def\natexlab#1{#1}\fi
\providecommand{\url}[1]{\texttt{#1}}
\providecommand{\href}[2]{#2}
\providecommand{\path}[1]{#1}
\providecommand{\DOIprefix}{doi:}
\providecommand{\ArXivprefix}{arXiv:}
\providecommand{\URLprefix}{URL: }
\providecommand{\Pubmedprefix}{pmid:}
\providecommand{\doi}[1]{\href{http://dx.doi.org/#1}{\path{#1}}}
\providecommand{\Pubmed}[1]{\href{pmid:#1}{\path{#1}}}
\providecommand{\bibinfo}[2]{#2}
\ifx\xfnm\relax \def\xfnm[#1]{\unskip,\space#1}\fi
\bibitem[{isi(2018)}]{isic2018}
, \bibinfo{year}{2018}.
\newblock \bibinfo{title}{{ISIC} 2018: Skin lesion analysis towards melanoma detection}.
\newblock \bibinfo{howpublished}{\url{https://challenge.isic-archive.com/data/\#2018}}.
\newblock \bibinfo{note}{Accessed: 2023-04-12}.
\bibitem[{Ayana et~al.(2023)Ayana, Dese, Dereje, Kebede, Barki, Amdissa, Husen, Mulugeta, Habtamu and Choe}]{diagnostics13020178}
\bibinfo{author}{Ayana, G.}, \bibinfo{author}{Dese, K.}, \bibinfo{author}{Dereje, Y.}, \bibinfo{author}{Kebede, Y.}, \bibinfo{author}{Barki, H.}, \bibinfo{author}{Amdissa, D.}, \bibinfo{author}{Husen, N.}, \bibinfo{author}{Mulugeta, F.}, \bibinfo{author}{Habtamu, B.}, \bibinfo{author}{Choe, S.W.}, \bibinfo{year}{2023}.
\newblock \bibinfo{title}{Vision-transformer-based transfer learning for mammogram classification}.
\newblock \bibinfo{journal}{Diagnostics} \bibinfo{volume}{13}.
\newblock \URLprefix \url{https://www.mdpi.com/2075-4418/13/2/178}, \DOIprefix\doi{10.3390/diagnostics13020178}.
\bibitem[{Bakker et~al.(2018)Bakker, Jacobs, Pondaag, Gelderblom, Nout, Dijkstra, Peul and Vleggeert-Lankamp}]{Bakker2018}
\bibinfo{author}{Bakker, S.H.}, \bibinfo{author}{Jacobs, W.C.H.}, \bibinfo{author}{Pondaag, W.}, \bibinfo{author}{Gelderblom, H.}, \bibinfo{author}{Nout, R.A.}, \bibinfo{author}{Dijkstra, P.D.S.}, \bibinfo{author}{Peul, W.C.}, \bibinfo{author}{Vleggeert-Lankamp, C.L.A.}, \bibinfo{year}{2018}.
\newblock \bibinfo{title}{Chordoma: a systematic review of the epidemiology and clinical prognostic factors predicting progression-free and overall survival}.
\newblock \bibinfo{journal}{European Spine Journal} \bibinfo{volume}{27}, \bibinfo{pages}{3043--3058}.
\newblock \URLprefix \url{https://doi.org/10.1007/s00586-018-5764-0}, \DOIprefix\doi{10.1007/s00586-018-5764-0}.
\bibitem[{Baldota et~al.(2021)Baldota, Sharma and Malathy}]{baldota2021deep}
\bibinfo{author}{Baldota, S.}, \bibinfo{author}{Sharma, S.}, \bibinfo{author}{Malathy, C.}, \bibinfo{year}{2021}.
\newblock \bibinfo{title}{Deep transfer learning for pancreatic cancer detection}, in: \bibinfo{booktitle}{2021 12th International Conference on Computing Communication and Networking Technologies (ICCCNT)}, \bibinfo{address}{Kharagpur, India}. pp. \bibinfo{pages}{1--7}.
\newblock \DOIprefix\doi{10.1109/ICCCNT51525.2021.9580000}.
\bibitem[{Baykul et~al.(2010)Baykul, Yilmaz, Aydin, Aydin, Aksoy and Yildirim}]{baykul2010early}
\bibinfo{author}{Baykul, T.}, \bibinfo{author}{Yilmaz, H.H.}, \bibinfo{author}{Aydin, U.}, \bibinfo{author}{Aydin, M.A.}, \bibinfo{author}{Aksoy, M.}, \bibinfo{author}{Yildirim, D.}, \bibinfo{year}{2010}.
\newblock \bibinfo{title}{Early diagnosis of oral cancer}.
\newblock \bibinfo{journal}{The Journal of International Medical Research} \bibinfo{volume}{38}, \bibinfo{pages}{737--749}.
\newblock \DOIprefix\doi{10.1177/147323001003800302}.
\bibitem[{Brinker et~al.(2019)Brinker, Hekler, Enk, Berking, Haferkamp, Hauschild, Weichenthal, Klode, Schadendorf, Holland-Letz, von Kalle, Fröhling, Schilling and Utikal}]{brinker2019deep}
\bibinfo{author}{Brinker, T.J.}, \bibinfo{author}{Hekler, A.}, \bibinfo{author}{Enk, A.H.}, \bibinfo{author}{Berking, C.}, \bibinfo{author}{Haferkamp, S.}, \bibinfo{author}{Hauschild, A.}, \bibinfo{author}{Weichenthal, M.}, \bibinfo{author}{Klode, J.}, \bibinfo{author}{Schadendorf, D.}, \bibinfo{author}{Holland-Letz, T.}, \bibinfo{author}{von Kalle, C.}, \bibinfo{author}{Fröhling, S.}, \bibinfo{author}{Schilling, B.}, \bibinfo{author}{Utikal, J.S.}, \bibinfo{year}{2019}.
\newblock \bibinfo{title}{Deep neural networks are superior to dermatologists in melanoma image classification}.
\newblock \bibinfo{journal}{European Journal of Cancer} \bibinfo{volume}{119}, \bibinfo{pages}{11--17}.
\bibitem[{Cancer.net(2022)}]{cancer_net_2022}
\bibinfo{author}{Cancer.net}, \bibinfo{year}{2022}.
\newblock \bibinfo{title}{Oral and oropharyngeal cancer: Statistics}.
\newblock \URLprefix \url{https://www.cancer.net/cancer-types/oral-and-oropharyngeal-cancer/statistics}. \bibinfo{note}{adapted from the American Cancer Society's (ACS) publication, Cancer Facts \& Figures 2022, the ACS website, the International Agency for Research on Cancer website, and the National Cancer Institute's Surveillance, Epidemiology, and End Results (SEER) Program. (All sources accessed February 2022.)}.
\bibitem[{Caron and Larochelle(2021)}]{vicreg}
\bibinfo{author}{Caron, M.}, \bibinfo{author}{Larochelle, H.}, \bibinfo{year}{2021}.
\newblock \bibinfo{title}{{VicReg}: Variance-invariance-covariance regularization}.
\newblock \bibinfo{howpublished}{GitHub repository}.
\newblock \bibinfo{note}{Https://github.com/facebookresearch/vicreg}.
\bibitem[{Chen et~al.(2020)Chen, Kornblith, Norouzi and Hinton}]{chen2020simple}
\bibinfo{author}{Chen, T.}, \bibinfo{author}{Kornblith, S.}, \bibinfo{author}{Norouzi, M.}, \bibinfo{author}{Hinton, G.}, \bibinfo{year}{2020}.
\newblock \bibinfo{title}{A simple framework for contrastive learning of visual representations}, in: \bibinfo{booktitle}{International Conference on Learning Representations (ICLR)}.
\bibitem[{Clark et~al.(2013)Clark, Vendt, Smith, Freymann, Kirby, Koppel, Moore, Phillips, Maffitt, Pringle, Tarbox and Prior}]{Clark2013}
\bibinfo{author}{Clark, K.}, \bibinfo{author}{Vendt, B.}, \bibinfo{author}{Smith, K.}, \bibinfo{author}{Freymann, J.}, \bibinfo{author}{Kirby, J.}, \bibinfo{author}{Koppel, P.}, \bibinfo{author}{Moore, S.}, \bibinfo{author}{Phillips, S.}, \bibinfo{author}{Maffitt, D.}, \bibinfo{author}{Pringle, M.}, \bibinfo{author}{Tarbox, L.}, \bibinfo{author}{Prior, F.}, \bibinfo{year}{2013}.
\newblock \bibinfo{title}{The cancer imaging archive (tcia): maintaining and operating a public information repository}.
\newblock \bibinfo{journal}{Journal of Digital Imaging} \bibinfo{volume}{26}, \bibinfo{pages}{1045--1057}.
\newblock \URLprefix \url{https://link.springer.com/article/10.1007/s10278-013-9622-7}, \DOIprefix\doi{10.1007/s10278-013-9622-7}.
\bibitem[{Deng et~al.(2009)Deng, Dong, Socher, Li, Li and Fei-Fei}]{Deng2009}
\bibinfo{author}{Deng, J.}, \bibinfo{author}{Dong, W.}, \bibinfo{author}{Socher, R.}, \bibinfo{author}{Li, L.J.}, \bibinfo{author}{Li, K.}, \bibinfo{author}{Fei-Fei, L.}, \bibinfo{year}{2009}.
\newblock \bibinfo{title}{{ImageNet: A Large-Scale Hierarchical Image Database}}, in: \bibinfo{booktitle}{2009 IEEE Conference on Computer Vision and Pattern Recognition}, pp. \bibinfo{pages}{248--255}.
\newblock \URLprefix \url{https://ieeexplore.ieee.org/document/5206848}, \DOIprefix\doi{10.1109/CVPR.2009.5206848}.
\bibitem[{{DermNet NZ}(n.d.)}]{dermnet}
\bibinfo{author}{{DermNet NZ}}, \bibinfo{year}{n.d.}
\newblock \bibinfo{title}{{DermNet NZ Image Repository}}.
\newblock \bibinfo{howpublished}{\url{https://dermnetnz.org/}}.
\newblock \bibinfo{note}{Accessed: 2024}.
\bibitem[{Dildar et~al.(2021)Dildar, Akram, Irfan, Khan, Ramzan, Mahmood, Alsaiari, Saeed, Alraddadi and Mahnashi}]{dildar2021skin}
\bibinfo{author}{Dildar, M.}, \bibinfo{author}{Akram, S.}, \bibinfo{author}{Irfan, M.}, \bibinfo{author}{Khan, H.U.}, \bibinfo{author}{Ramzan, M.}, \bibinfo{author}{Mahmood, A.R.}, \bibinfo{author}{Alsaiari, S.A.}, \bibinfo{author}{Saeed, A.H.M.}, \bibinfo{author}{Alraddadi, M.O.}, \bibinfo{author}{Mahnashi, M.H.}, \bibinfo{year}{2021}.
\newblock \bibinfo{title}{Skin cancer detection: A review using deep learning techniques}.
\newblock \bibinfo{journal}{International Journal of Environmental Research and Public Health} \bibinfo{volume}{18}, \bibinfo{pages}{5479}.
\bibitem[{Dosovitskiy et~al.(2020)Dosovitskiy, Beyer, Kolesnikov, Weissenborn, Zhai, Unterthiner, Dehghani, Minderer, Heigold, Gelly, Uszkoreit and Houlsby}]{dosovitskiy2020image}
\bibinfo{author}{Dosovitskiy, A.}, \bibinfo{author}{Beyer, L.}, \bibinfo{author}{Kolesnikov, A.}, \bibinfo{author}{Weissenborn, D.}, \bibinfo{author}{Zhai, X.}, \bibinfo{author}{Unterthiner, T.}, \bibinfo{author}{Dehghani, M.}, \bibinfo{author}{Minderer, M.}, \bibinfo{author}{Heigold, G.}, \bibinfo{author}{Gelly, S.}, \bibinfo{author}{Uszkoreit, J.}, \bibinfo{author}{Houlsby, N.}, \bibinfo{year}{2020}.
\newblock \bibinfo{title}{An image is worth 16x16 words: Transformers for image recognition at scale}.
\newblock \bibinfo{journal}{arXiv preprint arXiv:2010.11929} .
\bibitem[{Esteva et~al.(2017)Esteva, Kuprel, Novoa, Ko, Swetter, Blau and Thrun}]{esteva2017dermatologist}
\bibinfo{author}{Esteva, A.}, \bibinfo{author}{Kuprel, B.}, \bibinfo{author}{Novoa, R.A.}, \bibinfo{author}{Ko, J.}, \bibinfo{author}{Swetter, S.M.}, \bibinfo{author}{Blau, H.M.}, \bibinfo{author}{Thrun, S.}, \bibinfo{year}{2017}.
\newblock \bibinfo{title}{Dermatologist-level classification of skin cancer with deep neural networks}.
\newblock \bibinfo{journal}{Nature} \bibinfo{volume}{542}, \bibinfo{pages}{115--118}.
\newblock \URLprefix \url{https://doi-org.wwwproxy1.library.unsw.edu.au/10.1038/nature21056}, \DOIprefix\doi{10.1038/nature21056}.
\bibitem[{Fasilkebede(2023)}]{fasilkebede2023oral}
\bibinfo{author}{Fasilkebede, A.}, \bibinfo{year}{2023}.
\newblock \bibinfo{title}{{Oral Cancer Histopathology Dataset}}.
\newblock \bibinfo{howpublished}{Kaggle}.
\newblock \URLprefix \url{https://www.kaggle.com/datasets/ashenafifasilkebede/dataset}. \bibinfo{note}{accessed: 2024}.
\bibitem[{fast.ai(2021)}]{fastai_onecycle}
\bibinfo{author}{fast.ai}, \bibinfo{year}{2021}.
\newblock \bibinfo{title}{One cycle policy}.
\newblock \URLprefix \url{https://fastai1.fast.ai/callbacks.one_cycle.html}. \bibinfo{note}{accessed: 2023-03-29}.
\bibitem[{Ferlay et~al.(2021)Ferlay, Colombet, Soerjomataram, Parkin, Piñeros, Znaor and Bray}]{ferlay2021cancer}
\bibinfo{author}{Ferlay, J.}, \bibinfo{author}{Colombet, M.}, \bibinfo{author}{Soerjomataram, I.}, \bibinfo{author}{Parkin, D.M.}, \bibinfo{author}{Piñeros, M.}, \bibinfo{author}{Znaor, A.}, \bibinfo{author}{Bray, F.}, \bibinfo{year}{2021}.
\newblock \bibinfo{title}{Cancer statistics for the year 2020: An overview}.
\newblock \bibinfo{journal}{Int J Cancer} \bibinfo{note}{Online ahead of print}.
\bibitem[{Filella and Foj(2016)}]{filella2016prostate}
\bibinfo{author}{Filella, X.}, \bibinfo{author}{Foj, L.}, \bibinfo{year}{2016}.
\newblock \bibinfo{title}{Prostate cancer detection and prognosis: from prostate specific antigen (psa) to exosomal biomarkers}.
\newblock \bibinfo{journal}{International journal of molecular sciences} \bibinfo{volume}{17}, \bibinfo{pages}{1784}.
\bibitem[{Goel(2020)}]{dermnet_kaggle}
\bibinfo{author}{Goel, S.}, \bibinfo{year}{2020}.
\newblock \bibinfo{title}{{DermNet: Skin Disease Images Dataset (Watermarked version)}}.
\newblock \bibinfo{howpublished}{Kaggle}.
\newblock \URLprefix \url{https://www.kaggle.com/datasets/shubhamgoel27/dermnet}. \bibinfo{note}{accessed: 2024}.
\bibitem[{Goodfellow et~al.(2016)Goodfellow, Bengio and Courville}]{goodfellow2016deep}
\bibinfo{author}{Goodfellow, I.}, \bibinfo{author}{Bengio, Y.}, \bibinfo{author}{Courville, A.}, \bibinfo{year}{2016}.
\newblock \bibinfo{title}{Deep Learning}.
\newblock \bibinfo{publisher}{MIT Press}.
\newblock \URLprefix \url{http://www.deeplearningbook.org}.
\bibitem[{Grill et~al.(2020)Grill, Strub, Altché, Tallec, Richemond, Buchatskaya, Doersch, Pires, Guo, Azar, Piot, Kavukcuoglu, Munos and Valko}]{grill2020bootstrap}
\bibinfo{author}{Grill, J.B.}, \bibinfo{author}{Strub, F.}, \bibinfo{author}{Altché, F.}, \bibinfo{author}{Tallec, C.}, \bibinfo{author}{Richemond, P.H.}, \bibinfo{author}{Buchatskaya, E.}, \bibinfo{author}{Doersch, C.}, \bibinfo{author}{Pires, B.A.}, \bibinfo{author}{Guo, Z.D.}, \bibinfo{author}{Azar, M.G.}, \bibinfo{author}{Piot, B.}, \bibinfo{author}{Kavukcuoglu, K.}, \bibinfo{author}{Munos, R.}, \bibinfo{author}{Valko, M.}, \bibinfo{year}{2020}.
\newblock \bibinfo{title}{Bootstrap your own latent: A new approach to self-supervised learning}.
\newblock \bibinfo{journal}{Advances in Neural Information Processing Systems} \bibinfo{volume}{33}.
\bibitem[{Guo and Ashour(2018)}]{guo2018multiple}
\bibinfo{author}{Guo, Y.}, \bibinfo{author}{Ashour, A.S.}, \bibinfo{year}{2018}.
\newblock \bibinfo{title}{Multiple convolutional neural network for skin dermoscopic image classification}.
\newblock \bibinfo{journal}{arXiv preprint} \href{http://arxiv.org/abs/1807.08114}{\tt arXiv:1807.08114}.
\bibitem[{He et~al.(2016)He, Zhang, Ren and Sun}]{he2016deep}
\bibinfo{author}{He, K.}, \bibinfo{author}{Zhang, X.}, \bibinfo{author}{Ren, S.}, \bibinfo{author}{Sun, J.}, \bibinfo{year}{2016}.
\newblock \bibinfo{title}{Deep residual learning for image recognition}, in: \bibinfo{booktitle}{Proceedings of the IEEE conference on computer vision and pattern recognition}, pp. \bibinfo{pages}{770--778}.
\bibitem[{Howard and fastai team(2024)}]{fastai_lr_find}
\bibinfo{author}{Howard, J.}, \bibinfo{author}{fastai team}, \bibinfo{year}{2024}.
\newblock \bibinfo{title}{Learner.lr\_find function}.
\newblock \bibinfo{howpublished}{\textit{fastai Documentation}. Available at: \url{https://docs.fast.ai/callback.schedule.html\#learner.lr\_find}}.
\newblock \bibinfo{note}{Accessed: 21 October 2024}.
\bibitem[{Hua et~al.(2021)Hua, Wang, Xue, Ren, Wang and Zhao}]{hua2021feature}
\bibinfo{author}{Hua, T.}, \bibinfo{author}{Wang, W.}, \bibinfo{author}{Xue, Z.}, \bibinfo{author}{Ren, S.}, \bibinfo{author}{Wang, Y.}, \bibinfo{author}{Zhao, H.}, \bibinfo{year}{2021}.
\newblock \bibinfo{title}{On feature decorrelation in self-supervised learning}, in: \bibinfo{booktitle}{Proceedings of the IEEE/CVF International Conference on Computer Vision (ICCV)}.
\newblock \bibinfo{note}{Oral Presentation}.
\bibitem[{Huang et~al.(2023)Huang, Pareek, Jensen et~al.}]{Huang2023}
\bibinfo{author}{Huang, S.C.}, \bibinfo{author}{Pareek, A.}, \bibinfo{author}{Jensen, M.}, et~al., \bibinfo{year}{2023}.
\newblock \bibinfo{title}{Self-supervised learning for medical image classification: a systematic review and implementation guidelines}.
\newblock \bibinfo{journal}{npj Digital Medicine} \bibinfo{volume}{6}, \bibinfo{pages}{74}.
\newblock \DOIprefix\doi{10.1038/s41746-023-00811-0}.
\bibitem[{III et~al.(2011)III, McLennan, Bidaut, McNitt-Gray, Meyer, Reeves, Gamsu, Henschke, Hoffman, Kazerooni, MacMahon, Beek, Aberle, Yankaskas, Austin, Goldin, Prokop, Cody, Lynch, Mazzone, Fenton, van Ginneken, Lambin, Brown, Barnhart, Kalpathy-Cramer, Freymann, Kirby, Gavrielides, Kiciak and Bakis}]{Armato2011}
\bibinfo{author}{III, S.G.A.}, \bibinfo{author}{McLennan, G.}, \bibinfo{author}{Bidaut, L.}, \bibinfo{author}{McNitt-Gray, M.F.}, \bibinfo{author}{Meyer, C.R.}, \bibinfo{author}{Reeves, A.P.}, \bibinfo{author}{Gamsu, G.}, \bibinfo{author}{Henschke, C.}, \bibinfo{author}{Hoffman, E.A.}, \bibinfo{author}{Kazerooni, E.A.}, \bibinfo{author}{MacMahon, H.}, \bibinfo{author}{Beek, E.J.R.V.}, \bibinfo{author}{Aberle, D.R.}, \bibinfo{author}{Yankaskas, B.}, \bibinfo{author}{Austin, P.J.}, \bibinfo{author}{Goldin, J.}, \bibinfo{author}{Prokop, A.F.}, \bibinfo{author}{Cody, D.D.}, \bibinfo{author}{Lynch, D.A.}, \bibinfo{author}{Mazzone, J.C.}, \bibinfo{author}{Fenton, L.E.}, \bibinfo{author}{van Ginneken, B.}, \bibinfo{author}{Lambin, P.}, \bibinfo{author}{Brown, M.S.}, \bibinfo{author}{Barnhart, R.S.}, \bibinfo{author}{Kalpathy-Cramer}, \bibinfo{author}{Freymann, J.E.}, \bibinfo{author}{Kirby, J.S.}, \bibinfo{author}{Gavrielides, M.A.}, \bibinfo{author}{Kiciak, P.B.}, \bibinfo{author}{Bakis, C.E.},
  \bibinfo{year}{2011}.
\newblock \bibinfo{title}{The lung image database consortium (lidc) and image database resource initiative (idri): a completed reference database of lung nodules on ct scans}.
\newblock \bibinfo{journal}{Medical Physics} \bibinfo{volume}{38}, \bibinfo{pages}{915--931}.
\newblock \URLprefix \url{https://doi.org/10.1118/1.3528204}, \DOIprefix\doi{10.1118/1.3528204}.
\bibitem[{{International Skin Imaging Collaboration}(2023)}]{ISIC2023}
\bibinfo{author}{{International Skin Imaging Collaboration}}, \bibinfo{year}{2023}.
\newblock \bibinfo{title}{{ISIC} archive: A comprehensive resource for skin imaging data}.
\newblock \URLprefix \url{https://www.isic-archive.com}. \bibinfo{note}{accessed: 2023-04-20}.
\bibitem[{Jabbar et~al.(2022)Jabbar, Yan, Xu, Ur~Rehman and Jabbar}]{jabbar2022transfer}
\bibinfo{author}{Jabbar, M.K.}, \bibinfo{author}{Yan, J.}, \bibinfo{author}{Xu, H.}, \bibinfo{author}{Ur~Rehman, Z.}, \bibinfo{author}{Jabbar, A.}, \bibinfo{year}{2022}.
\newblock \bibinfo{title}{Transfer learning-based model for diabetic retinopathy diagnosis using retinal images}.
\newblock \bibinfo{journal}{Brain Sciences} \bibinfo{volume}{12}, \bibinfo{pages}{535}.
\newblock \DOIprefix\doi{10.3390/brainsci12050535}.
\bibitem[{Jaiswal et~al.(2020)Jaiswal, Babu, Zadeh, Banerjee and Makedon}]{jaiswal2020survey}
\bibinfo{author}{Jaiswal, A.}, \bibinfo{author}{Babu, A.R.}, \bibinfo{author}{Zadeh, M.Z.}, \bibinfo{author}{Banerjee, D.}, \bibinfo{author}{Makedon, F.}, \bibinfo{year}{2020}.
\newblock \bibinfo{title}{A survey on contrastive self-supervised learning}.
\newblock \bibinfo{journal}{arXiv preprint arXiv:2011.00362} .
\bibitem[{Jarvis and Williams(2016)}]{jarvis2016irreproducibility}
\bibinfo{author}{Jarvis, M.F.}, \bibinfo{author}{Williams, M.}, \bibinfo{year}{2016}.
\newblock \bibinfo{title}{Irreproducibility in preclinical biomedical research: perceptions, uncertainties, and knowledge gaps}.
\newblock \bibinfo{journal}{Trends in pharmacological sciences} \bibinfo{volume}{37}, \bibinfo{pages}{290--302}.
\bibitem[{Jeschke et~al.(2012)Jeschke, Van~Neste, Gl{\"o}ckner, Dhir, Calmon, Deregowski et~al.}]{jeschke2012biomarkers}
\bibinfo{author}{Jeschke, J.}, \bibinfo{author}{Van~Neste, L.}, \bibinfo{author}{Gl{\"o}ckner, S.C.}, \bibinfo{author}{Dhir, M.}, \bibinfo{author}{Calmon, M.F.}, \bibinfo{author}{Deregowski, V.}, et~al., \bibinfo{year}{2012}.
\newblock \bibinfo{title}{Biomarkers for detection and prognosis of breast cancer identified by a functional hypermethylome screen}.
\newblock \bibinfo{journal}{Epigenetics} \bibinfo{volume}{7}, \bibinfo{pages}{701--709}.
\bibitem[{Jing et~al.(2022)Jing, Vincent, LeCun and Tian}]{jing2022understanding}
\bibinfo{author}{Jing, L.}, \bibinfo{author}{Vincent, P.}, \bibinfo{author}{LeCun, Y.}, \bibinfo{author}{Tian, Y.}, \bibinfo{year}{2022}.
\newblock \bibinfo{title}{Understanding dimensional collapse in contrastive self-supervised learning}, in: \bibinfo{booktitle}{International Conference on Learning Representations}.
\bibitem[{Kamisawa et~al.(2016)Kamisawa, Wood, Itoi and Takaori}]{kamisawa2016pancreatic}
\bibinfo{author}{Kamisawa, T.}, \bibinfo{author}{Wood, L.D.}, \bibinfo{author}{Itoi, T.}, \bibinfo{author}{Takaori, K.}, \bibinfo{year}{2016}.
\newblock \bibinfo{title}{Pancreatic cancer}.
\newblock \bibinfo{journal}{The Lancet} \bibinfo{volume}{388}, \bibinfo{pages}{73--85}.
\bibitem[{Kaplan et~al.(2020)Kaplan, McCandlish, Henighan, Brown, Chess, Child, Gray, Radford, Wu and Amodei}]{kaplan2020scalinglawsneurallanguage}
\bibinfo{author}{Kaplan, J.}, \bibinfo{author}{McCandlish, S.}, \bibinfo{author}{Henighan, T.}, \bibinfo{author}{Brown, T.B.}, \bibinfo{author}{Chess, B.}, \bibinfo{author}{Child, R.}, \bibinfo{author}{Gray, S.}, \bibinfo{author}{Radford, A.}, \bibinfo{author}{Wu, J.}, \bibinfo{author}{Amodei, D.}, \bibinfo{year}{2020}.
\newblock \bibinfo{title}{Scaling laws for neural language models}.
\newblock \URLprefix \url{https://arxiv.org/abs/2001.08361}, \href{http://arxiv.org/abs/2001.08361}{\tt arXiv:2001.08361}.
\bibitem[{Kaur et~al.(2020)Kaur, Singla, Nkenyereye, Jha, Prashar, Joshi, El-Sappagh, Islam and Islam}]{kaur2020medical}
\bibinfo{author}{Kaur, S.}, \bibinfo{author}{Singla, J.}, \bibinfo{author}{Nkenyereye, L.}, \bibinfo{author}{Jha, S.}, \bibinfo{author}{Prashar, D.}, \bibinfo{author}{Joshi, G.P.}, \bibinfo{author}{El-Sappagh, S.}, \bibinfo{author}{Islam, M.S.}, \bibinfo{author}{Islam, S.R.}, \bibinfo{year}{2020}.
\newblock \bibinfo{title}{Medical diagnostic systems using artificial intelligence (ai) algorithms: Principles and perspectives}.
\newblock \bibinfo{journal}{IEEE Access} \bibinfo{volume}{8}, \bibinfo{pages}{228049--228069}.
\bibitem[{Kazarian et~al.(2017)Kazarian, Blyuss, Metodieva, Gentry-Maharaj, Ryan, Kiseleva et~al.}]{kazarian2017testing}
\bibinfo{author}{Kazarian, A.}, \bibinfo{author}{Blyuss, O.}, \bibinfo{author}{Metodieva, G.}, \bibinfo{author}{Gentry-Maharaj, A.}, \bibinfo{author}{Ryan, A.}, \bibinfo{author}{Kiseleva, E.M.}, et~al., \bibinfo{year}{2017}.
\newblock \bibinfo{title}{Testing breast cancer serum biomarkers for early detection and prognosis in pre-diagnosis samples}.
\newblock \bibinfo{journal}{British journal of cancer} \bibinfo{volume}{116}, \bibinfo{pages}{501--508}.
\bibitem[{Khan et~al.(2023)Khan, Akbar, Mehmood, Shahid, Munir, Ilyas, Asif and Zheng}]{khan2023transfer}
\bibinfo{author}{Khan, R.}, \bibinfo{author}{Akbar, S.}, \bibinfo{author}{Mehmood, A.}, \bibinfo{author}{Shahid, F.}, \bibinfo{author}{Munir, K.}, \bibinfo{author}{Ilyas, N.}, \bibinfo{author}{Asif, M.}, \bibinfo{author}{Zheng, Z.}, \bibinfo{year}{2023}.
\newblock \bibinfo{title}{A transfer learning approach for multiclass classification of {Alzheimer}'s disease using {MRI} images}.
\newblock \bibinfo{journal}{Frontiers in Neuroscience} \bibinfo{volume}{16}, \bibinfo{pages}{1050777}.
\newblock \DOIprefix\doi{10.3389/fnins.2022.1050777}.
\bibitem[{Kim et~al.(2022)Kim, Cosa-Linan, Santhanam, Jannesari, Maros and Ganslandt}]{kim2022transfer}
\bibinfo{author}{Kim, H.E.}, \bibinfo{author}{Cosa-Linan, A.}, \bibinfo{author}{Santhanam, N.}, \bibinfo{author}{Jannesari, M.}, \bibinfo{author}{Maros, M.E.}, \bibinfo{author}{Ganslandt, T.}, \bibinfo{year}{2022}.
\newblock \bibinfo{title}{Transfer learning for medical image classification: a literature review}.
\newblock \bibinfo{journal}{BMC Medical Imaging} \bibinfo{volume}{22}, \bibinfo{pages}{69}.
\bibitem[{Kingma and Ba(2014)}]{kingma2014adam}
\bibinfo{author}{Kingma, D.P.}, \bibinfo{author}{Ba, J.}, \bibinfo{year}{2014}.
\newblock \bibinfo{title}{Adam: A method for stochastic optimization}.
\newblock \bibinfo{journal}{arXiv preprint arXiv:1412.6980} .
\bibitem[{Kwong and Mazaheri(2021)}]{kwong2021survey}
\bibinfo{author}{Kwong, T.}, \bibinfo{author}{Mazaheri, S.}, \bibinfo{year}{2021}.
\newblock \bibinfo{title}{A survey on deep learning approaches for breast cancer diagnosis}.
\newblock \bibinfo{journal}{arXiv preprint} \href{http://arxiv.org/abs/2109.08853}{\tt arXiv:2109.08853}.
\bibitem[{LeCun and Misra(2021)}]{lecun2021self}
\bibinfo{author}{LeCun, Y.}, \bibinfo{author}{Misra, I.}, \bibinfo{year}{2021}.
\newblock \bibinfo{title}{Self-supervised learning: The dark matter of intelligence}.
\newblock \bibinfo{howpublished}{Facebook AI Blog}.
\bibitem[{Litjens et~al.(2017)Litjens, Kooi, Bejnordi, Setio, Ciompi, Ghafoorian, van~der Laak, van Ginneken and Sánchez}]{Litjens2017}
\bibinfo{author}{Litjens, G.}, \bibinfo{author}{Kooi, T.}, \bibinfo{author}{Bejnordi, B.E.}, \bibinfo{author}{Setio, A.A.A.}, \bibinfo{author}{Ciompi, F.}, \bibinfo{author}{Ghafoorian, M.}, \bibinfo{author}{van~der Laak, J.A.}, \bibinfo{author}{van Ginneken, B.}, \bibinfo{author}{Sánchez, C.I.}, \bibinfo{year}{2017}.
\newblock \bibinfo{title}{A survey on deep learning in medical image analysis}.
\newblock \bibinfo{journal}{Medical Image Analysis} \bibinfo{volume}{42}, \bibinfo{pages}{60--88}.
\newblock \URLprefix \url{https://www.sciencedirect.com/science/article/pii/S1361841517301135}, \DOIprefix\doi{10.1016/j.media.2017.07.005}.
\bibitem[{Lopez et~al.(2017)Lopez, Giro-i Nieto, Burdick and Marques}]{lopez2017skin}
\bibinfo{author}{Lopez, A.R.}, \bibinfo{author}{Giro-i Nieto, X.}, \bibinfo{author}{Burdick, J.}, \bibinfo{author}{Marques, O.}, \bibinfo{year}{2017}.
\newblock \bibinfo{title}{Skin lesion classification from dermoscopic images using deep learning techniques}, in: \bibinfo{booktitle}{2017 13th IASTED international conference on biomedical engineering (BioMed)}, \bibinfo{organization}{IEEE}. pp. \bibinfo{pages}{49--54}.
\bibitem[{Mahbod et~al.(2019)Mahbod, Schaefer, Wang, Ecker and Ellinge}]{mahbod2019skin}
\bibinfo{author}{Mahbod, A.}, \bibinfo{author}{Schaefer, G.}, \bibinfo{author}{Wang, C.}, \bibinfo{author}{Ecker, R.}, \bibinfo{author}{Ellinge, I.}, \bibinfo{year}{2019}.
\newblock \bibinfo{title}{Skin lesion classification using hybrid deep neural networks}, in: \bibinfo{booktitle}{ICASSP 2019-2019 IEEE International Conference on Acoustics, Speech and Signal Processing (ICASSP)}, \bibinfo{organization}{IEEE}. pp. \bibinfo{pages}{1229--1233}.
\bibitem[{Majtner et~al.(2018)Majtner, Bajić, Yildirim, Hardeberg, Lindblad and Sladoje}]{majtner2018ensemble}
\bibinfo{author}{Majtner, T.}, \bibinfo{author}{Bajić, B.}, \bibinfo{author}{Yildirim, S.}, \bibinfo{author}{Hardeberg, J.Y.}, \bibinfo{author}{Lindblad, J.}, \bibinfo{author}{Sladoje, N.}, \bibinfo{year}{2018}.
\newblock \bibinfo{title}{Ensemble of convolutional neural networks for dermoscopic images classification}.
\newblock \bibinfo{journal}{arXiv preprint} \href{http://arxiv.org/abs/1808.05071}{\tt arXiv:1808.05071}.
\bibitem[{Maron et~al.(2019)Maron, Weichenthal, Utikal, Hekler, Berking, Hauschild, Enk, Haferkamp, Klode, Schadendorf, Jansen, Holland-Letz, Schilling, {von Kalle}, Fröhling, Gaiser, Hartmann, Gesierich, Kähler, Wehkamp, Karoglan, Bär, Brinker, Schmitt, Peitsch, Hoffmann, Becker, Drusio, Jansen, Klode, Lodde, Sammet, Schadendorf, Sondermann, Ugurel, Zader, Enk, Salzmann, Schäfer, Schäkel, Winkler, Wölbing, Asper, Bohne, Brown, Burba, Deffaa, Dietrich, Dietrich, Drerup, Egberts, Erkens, Greven, Harde, Jost, Kaeding, Kosova, Lischner, Maagk, Messinger, Metzner, Motamedi, Rosenthal, Seidl, Stemmermann, Torz, Velez, Haiduk, Alter, Bär, Bergenthal, Gerlach, Holtorf, Karoglan, Kindermann, Kraas, Felcht, Gaiser, Klemke, Kurzen, Leibing, Müller, Reinhard, Utikal, Winter, Berking, Eicher, Hartmann, Heppt, Kilian, Krammer, Lill, Niesert, Oppel, Sattler, Senner, Wallmichrath, Wolff, Giner, Glutsch, Kerstan, Presser, Schrüfer, Schummer, Stolze, Weber, Drexler, Haferkamp, Mickler, Stauner and
  Thiem}]{MARON201957}
\bibinfo{author}{Maron, R.C.}, \bibinfo{author}{Weichenthal, M.}, \bibinfo{author}{Utikal, J.S.}, \bibinfo{author}{Hekler, A.}, \bibinfo{author}{Berking, C.}, \bibinfo{author}{Hauschild, A.}, \bibinfo{author}{Enk, A.H.}, \bibinfo{author}{Haferkamp, S.}, \bibinfo{author}{Klode, J.}, \bibinfo{author}{Schadendorf, D.}, \bibinfo{author}{Jansen, P.}, \bibinfo{author}{Holland-Letz, T.}, \bibinfo{author}{Schilling, B.}, \bibinfo{author}{{von Kalle}, C.}, \bibinfo{author}{Fröhling, S.}, \bibinfo{author}{Gaiser, M.R.}, \bibinfo{author}{Hartmann, D.}, \bibinfo{author}{Gesierich, A.}, \bibinfo{author}{Kähler, K.C.}, \bibinfo{author}{Wehkamp, U.}, \bibinfo{author}{Karoglan, A.}, \bibinfo{author}{Bär, C.}, \bibinfo{author}{Brinker, T.J.}, \bibinfo{author}{Schmitt, L.}, \bibinfo{author}{Peitsch, W.K.}, \bibinfo{author}{Hoffmann, F.}, \bibinfo{author}{Becker, J.C.}, \bibinfo{author}{Drusio, C.}, \bibinfo{author}{Jansen, P.}, \bibinfo{author}{Klode, J.}, \bibinfo{author}{Lodde, G.}, \bibinfo{author}{Sammet, S.},
  \bibinfo{author}{Schadendorf, D.}, \bibinfo{author}{Sondermann, W.}, \bibinfo{author}{Ugurel, S.}, \bibinfo{author}{Zader, J.}, \bibinfo{author}{Enk, A.}, \bibinfo{author}{Salzmann, M.}, \bibinfo{author}{Schäfer, S.}, \bibinfo{author}{Schäkel, K.}, \bibinfo{author}{Winkler, J.}, \bibinfo{author}{Wölbing, P.}, \bibinfo{author}{Asper, H.}, \bibinfo{author}{Bohne, A.S.}, \bibinfo{author}{Brown, V.}, \bibinfo{author}{Burba, B.}, \bibinfo{author}{Deffaa, S.}, \bibinfo{author}{Dietrich, C.}, \bibinfo{author}{Dietrich, M.}, \bibinfo{author}{Drerup, K.A.}, \bibinfo{author}{Egberts, F.}, \bibinfo{author}{Erkens, A.S.}, \bibinfo{author}{Greven, S.}, \bibinfo{author}{Harde, V.}, \bibinfo{author}{Jost, M.}, \bibinfo{author}{Kaeding, M.}, \bibinfo{author}{Kosova, K.}, \bibinfo{author}{Lischner, S.}, \bibinfo{author}{Maagk, M.}, \bibinfo{author}{Messinger, A.L.}, \bibinfo{author}{Metzner, M.}, \bibinfo{author}{Motamedi, R.}, \bibinfo{author}{Rosenthal, A.C.}, \bibinfo{author}{Seidl, U.}, \bibinfo{author}{Stemmermann,
  J.}, \bibinfo{author}{Torz, K.}, \bibinfo{author}{Velez, J.G.}, \bibinfo{author}{Haiduk, J.}, \bibinfo{author}{Alter, M.}, \bibinfo{author}{Bär, C.}, \bibinfo{author}{Bergenthal, P.}, \bibinfo{author}{Gerlach, A.}, \bibinfo{author}{Holtorf, C.}, \bibinfo{author}{Karoglan, A.}, \bibinfo{author}{Kindermann, S.}, \bibinfo{author}{Kraas, L.}, \bibinfo{author}{Felcht, M.}, \bibinfo{author}{Gaiser, M.R.}, \bibinfo{author}{Klemke, C.D.}, \bibinfo{author}{Kurzen, H.}, \bibinfo{author}{Leibing, T.}, \bibinfo{author}{Müller, V.}, \bibinfo{author}{Reinhard, R.R.}, \bibinfo{author}{Utikal, J.}, \bibinfo{author}{Winter, F.}, \bibinfo{author}{Berking, C.}, \bibinfo{author}{Eicher, L.}, \bibinfo{author}{Hartmann, D.}, \bibinfo{author}{Heppt, M.}, \bibinfo{author}{Kilian, K.}, \bibinfo{author}{Krammer, S.}, \bibinfo{author}{Lill, D.}, \bibinfo{author}{Niesert, A.C.}, \bibinfo{author}{Oppel, E.}, \bibinfo{author}{Sattler, E.}, \bibinfo{author}{Senner, S.}, \bibinfo{author}{Wallmichrath, J.}, \bibinfo{author}{Wolff, H.},
  \bibinfo{author}{Giner, T.}, \bibinfo{author}{Glutsch, V.}, \bibinfo{author}{Kerstan, A.}, \bibinfo{author}{Presser, D.}, \bibinfo{author}{Schrüfer, P.}, \bibinfo{author}{Schummer, P.}, \bibinfo{author}{Stolze, I.}, \bibinfo{author}{Weber, J.}, \bibinfo{author}{Drexler, K.}, \bibinfo{author}{Haferkamp, S.}, \bibinfo{author}{Mickler, M.}, \bibinfo{author}{Stauner, C.T.}, \bibinfo{author}{Thiem, A.}, \bibinfo{year}{2019}.
\newblock \bibinfo{title}{Systematic outperformance of 112 dermatologists in multiclass skin cancer image classification by convolutional neural networks}.
\newblock \bibinfo{journal}{European Journal of Cancer} \bibinfo{volume}{119}, \bibinfo{pages}{57--65}.
\newblock \URLprefix \url{https://www.sciencedirect.com/science/article/pii/S0959804919303818}, \DOIprefix\doi{https://doi.org/10.1016/j.ejca.2019.06.013}.
\bibitem[{Minaee et~al.(2024)Minaee, Mikolov, Nikzad, Chenaghlu, Socher, Amatriain and Gao}]{minaee2024largelanguagemodelssurvey}
\bibinfo{author}{Minaee, S.}, \bibinfo{author}{Mikolov, T.}, \bibinfo{author}{Nikzad, N.}, \bibinfo{author}{Chenaghlu, M.}, \bibinfo{author}{Socher, R.}, \bibinfo{author}{Amatriain, X.}, \bibinfo{author}{Gao, J.}, \bibinfo{year}{2024}.
\newblock \bibinfo{title}{Large language models: A survey}.
\newblock \URLprefix \url{https://arxiv.org/abs/2402.06196}, \href{http://arxiv.org/abs/2402.06196}{\tt arXiv:2402.06196}.
\bibitem[{Misra and van~der Maaten(2019)}]{misra2019self}
\bibinfo{author}{Misra, I.}, \bibinfo{author}{van~der Maaten, L.}, \bibinfo{year}{2019}.
\newblock \bibinfo{title}{Self-supervised learning of pretext-invariant representations}.
\newblock \bibinfo{journal}{arXiv preprint arXiv:1912.01991} .
\bibitem[{Pacheco et~al.(2020)Pacheco, Lima, Salom{\~a}o, Krohling, Biral, de~Angelo, Alves, Esgario, Simora, Castro, Rodrigues, Frasson, Krohling, Knidel, Santos, do~Esp{\'\i}rito~Santo, Macedo, Canuto and de~Barros}]{pacheco2020pad}
\bibinfo{author}{Pacheco, A.G.C.}, \bibinfo{author}{Lima, G.R.}, \bibinfo{author}{Salom{\~a}o, A.S.}, \bibinfo{author}{Krohling, B.}, \bibinfo{author}{Biral, I.P.}, \bibinfo{author}{de~Angelo, G.G.}, \bibinfo{author}{Alves, F.C.R.J.}, \bibinfo{author}{Esgario, J.G.M.}, \bibinfo{author}{Simora, A.C.}, \bibinfo{author}{Castro, P.B.C.}, \bibinfo{author}{Rodrigues, F.B.}, \bibinfo{author}{Frasson, P.H.L.}, \bibinfo{author}{Krohling, R.A.}, \bibinfo{author}{Knidel, H.}, \bibinfo{author}{Santos, M.C.S.}, \bibinfo{author}{do~Esp{\'\i}rito~Santo, R.B.}, \bibinfo{author}{Macedo, T.L.S.G.}, \bibinfo{author}{Canuto, T.R.P.}, \bibinfo{author}{de~Barros, L.F.S.}, \bibinfo{year}{2020}.
\newblock \bibinfo{title}{{PAD-UFES-20: A skin lesion dataset composed of patient data and clinical images collected from smartphones}}.
\newblock \bibinfo{journal}{Data in Brief} \bibinfo{volume}{32}, \bibinfo{pages}{106221}.
\newblock \DOIprefix\doi{10.1016/j.dib.2020.106221}. \bibinfo{note}{pMID: 32939378; PMCID: PMC7479321}.
\bibitem[{PyTorch(2021)}]{torchvision_resnet50}
\bibinfo{author}{PyTorch}, \bibinfo{year}{2021}.
\newblock \bibinfo{title}{torchvision.models.resnet50}.
\newblock \URLprefix \url{https://pytorch.org/vision/main/models/generated/torchvision.models.resnet50.html}. \bibinfo{note}{accessed: 2023-03-29}.
\bibitem[{Raghu et~al.(2020)Raghu, Raghu, Bengio and Vinyals}]{raghu2020rapid}
\bibinfo{author}{Raghu, A.}, \bibinfo{author}{Raghu, M.}, \bibinfo{author}{Bengio, S.}, \bibinfo{author}{Vinyals, O.}, \bibinfo{year}{2020}.
\newblock \bibinfo{title}{Rapid learning or feature reuse? towards understanding the effectiveness of maml}, in: \bibinfo{booktitle}{Proceedings of the International Conference on Learning Representations (ICLR)}, \bibinfo{organization}{ICLR}.
\newblock \URLprefix \url{https://arxiv.org/abs/1909.09157}. \bibinfo{note}{\textit{Published as a conference paper at ICLR 2020}}.
\bibitem[{Research(2021)}]{barlowtwins}
\bibinfo{author}{Research, F.}, \bibinfo{year}{2021}.
\newblock \bibinfo{title}{Barlow twins: {S}elf-{S}upervised {L}earning via {R}edundancy {R}eduction}.
\newblock \URLprefix \url{https://github.com/facebookresearch/barlowtwins}. \bibinfo{note}{accessed: 2023-03-29}.
\bibitem[{Rigel et~al.(2010)Rigel, Russak and Friedman}]{rigel2010evolution}
\bibinfo{author}{Rigel, D.S.}, \bibinfo{author}{Russak, J.}, \bibinfo{author}{Friedman, R.}, \bibinfo{year}{2010}.
\newblock \bibinfo{title}{The evolution of melanoma diagnosis: 25 years beyond the abcds}.
\newblock \bibinfo{journal}{CA: A Cancer Journal for Clinicians} \bibinfo{volume}{60}, \bibinfo{pages}{301--316}.
\bibitem[{Russakovsky et~al.(2015)Russakovsky, Deng, Su, Krause, Satheesh, Ma, Huang, Karpathy, Khosla, Bernstein, Berg and Fei-Fei}]{russakovsky2015ImageNet}
\bibinfo{author}{Russakovsky, O.}, \bibinfo{author}{Deng, J.}, \bibinfo{author}{Su, H.}, \bibinfo{author}{Krause, J.}, \bibinfo{author}{Satheesh, S.}, \bibinfo{author}{Ma, S.}, \bibinfo{author}{Huang, Z.}, \bibinfo{author}{Karpathy, A.}, \bibinfo{author}{Khosla, A.}, \bibinfo{author}{Bernstein, M.}, \bibinfo{author}{Berg, A.C.}, \bibinfo{author}{Fei-Fei, L.}, \bibinfo{year}{2015}.
\newblock \bibinfo{title}{Imagenet large scale visual recognition challenge}.
\newblock \bibinfo{journal}{International Journal of Computer Vision} \bibinfo{volume}{115}, \bibinfo{pages}{211--252}.
\newblock \DOIprefix\doi{10.1007/s11263-015-0816-y}.
\bibitem[{Sengupta et~al.(2022)Sengupta, Sarode, Sarode and Ghone}]{sengupta2022scarcity}
\bibinfo{author}{Sengupta, N.}, \bibinfo{author}{Sarode, S.C.}, \bibinfo{author}{Sarode, G.S.}, \bibinfo{author}{Ghone, U.}, \bibinfo{year}{2022}.
\newblock \bibinfo{title}{Scarcity of publicly available oral cancer image datasets for machine learning research}.
\newblock \bibinfo{journal}{Oral Oncology} \bibinfo{volume}{126}, \bibinfo{pages}{105737}.
\bibitem[{Shanmugam et~al.(2021)Shanmugam, Blalock, Balakrishnan and Guttag}]{9710313}
\bibinfo{author}{Shanmugam, D.}, \bibinfo{author}{Blalock, D.}, \bibinfo{author}{Balakrishnan, G.}, \bibinfo{author}{Guttag, J.}, \bibinfo{year}{2021}.
\newblock \bibinfo{title}{Better aggregation in test-time augmentation}, in: \bibinfo{booktitle}{2021 IEEE/CVF International Conference on Computer Vision (ICCV)}, \bibinfo{publisher}{IEEE Computer Society}. pp. \bibinfo{pages}{1194--1203}.
\newblock \URLprefix \url{https://doi.ieeecomputersociety.org/10.1109/ICCV48922.2021.00125}, \DOIprefix\doi{10.1109/ICCV48922.2021.00125}.
\bibitem[{Shiffrin et~al.(2018)Shiffrin, B{\"o}rner and Stigler}]{shiffrin2018scientific}
\bibinfo{author}{Shiffrin, R.M.}, \bibinfo{author}{B{\"o}rner, K.}, \bibinfo{author}{Stigler, S.M.}, \bibinfo{year}{2018}.
\newblock \bibinfo{title}{Scientific progress despite irreproducibility: A seeming paradox}.
\newblock \bibinfo{journal}{Proceedings of the National Academy of Sciences} \bibinfo{volume}{115}, \bibinfo{pages}{2632--2639}.
\bibitem[{Simonyan and Zisserman(2014)}]{simonyan2014very}
\bibinfo{author}{Simonyan, K.}, \bibinfo{author}{Zisserman, A.}, \bibinfo{year}{2014}.
\newblock \bibinfo{title}{Very deep convolutional networks for large-scale image recognition}.
\newblock \bibinfo{journal}{arXiv preprint arXiv:1409.1556} .
\bibitem[{Smith(2018)}]{smith2018disciplined}
\bibinfo{author}{Smith, L.N.}, \bibinfo{year}{2018}.
\newblock \bibinfo{title}{A disciplined approach to neural network hyper-parameters: Part 1 -- learning rate, batch size, momentum, and weight decay}.
\newblock \bibinfo{journal}{arXiv preprint arXiv:1803.09820} \href{http://arxiv.org/abs/1803.09820}{\tt arXiv:1803.09820}.
\bibitem[{Smith and Topin(2017)}]{smith2017super}
\bibinfo{author}{Smith, L.N.}, \bibinfo{author}{Topin, N.}, \bibinfo{year}{2017}.
\newblock \bibinfo{title}{Super-convergence: Very fast training of neural networks using large learning rates}.
\newblock \bibinfo{journal}{arXiv preprint arXiv:1708.07120} \href{http://arxiv.org/abs/1708.07120}{\tt arXiv:1708.07120}.
\bibitem[{Song et~al.(2021)Song, Li, Sunny, Gurushanth, Mendonca, Mukhia, Patrick, Gurudath, Raghavan, Tsusennaro, Leivon, Kolur, Shetty, Bushan, Ramesh, Peterson, Pillai, Wilder-Smith, Sigamani, Suresh, Kuriakose, Birur and Liang}]{song2021classification}
\bibinfo{author}{Song, B.}, \bibinfo{author}{Li, S.}, \bibinfo{author}{Sunny, S.}, \bibinfo{author}{Gurushanth, K.}, \bibinfo{author}{Mendonca, P.}, \bibinfo{author}{Mukhia, N.}, \bibinfo{author}{Patrick, S.}, \bibinfo{author}{Gurudath, S.}, \bibinfo{author}{Raghavan, S.}, \bibinfo{author}{Tsusennaro, I.}, \bibinfo{author}{Leivon, S.T.}, \bibinfo{author}{Kolur, T.}, \bibinfo{author}{Shetty, V.}, \bibinfo{author}{Bushan, V.}, \bibinfo{author}{Ramesh, R.}, \bibinfo{author}{Peterson, T.}, \bibinfo{author}{Pillai, V.}, \bibinfo{author}{Wilder-Smith, P.}, \bibinfo{author}{Sigamani, A.}, \bibinfo{author}{Suresh, A.}, \bibinfo{author}{Kuriakose, M.A.}, \bibinfo{author}{Birur, P.}, \bibinfo{author}{Liang, R.}, \bibinfo{year}{2021}.
\newblock \bibinfo{title}{Classification of imbalanced oral cancer image data from high-risk population}.
\newblock \bibinfo{journal}{J Biomed Opt} \bibinfo{volume}{26}, \bibinfo{pages}{105001}.
\bibitem[{Song et~al.(2023)Song, Su, Wang, Qiang, Zheng and Sun}]{song2023towards}
\bibinfo{author}{Song, Z.}, \bibinfo{author}{Su, X.}, \bibinfo{author}{Wang, J.}, \bibinfo{author}{Qiang, W.}, \bibinfo{author}{Zheng, C.}, \bibinfo{author}{Sun, F.}, \bibinfo{year}{2023}.
\newblock \bibinfo{title}{Towards the sparseness of projection head in self-supervised learning}.
\newblock \bibinfo{journal}{arXiv preprint arXiv:2303.12962} .
\bibitem[{Syed et~al.(2020)Syed, Sleeman~IV, Nalluri, Kapoor, Hagan, Palta and Ghosh}]{syed2020artificial}
\bibinfo{author}{Syed, K.}, \bibinfo{author}{Sleeman~IV, W.C.}, \bibinfo{author}{Nalluri, J.J.}, \bibinfo{author}{Kapoor, R.}, \bibinfo{author}{Hagan, M.}, \bibinfo{author}{Palta, J.}, \bibinfo{author}{Ghosh, P.}, \bibinfo{year}{2020}.
\newblock \bibinfo{title}{Artificial intelligence methods in computer-aided diagnostic tools and decision support analytics for clinical informatics}, in: \bibinfo{booktitle}{Artificial Intelligence in Precision Health}. \bibinfo{publisher}{Elsevier}, pp. \bibinfo{pages}{31--59}.
\bibitem[{Szolovits et~al.(1988)Szolovits, Patil and Schwartz}]{szolovits1988artificial}
\bibinfo{author}{Szolovits, P.}, \bibinfo{author}{Patil, R.S.}, \bibinfo{author}{Schwartz, W.B.}, \bibinfo{year}{1988}.
\newblock \bibinfo{title}{Artificial intelligence in medical diagnosis}.
\newblock \bibinfo{journal}{Annals of internal medicine} \bibinfo{volume}{108}, \bibinfo{pages}{80--87}.
\bibitem[{Tan et~al.(2018)Tan, Sun, Kong, Zhang, Yang and Liu}]{tan2018survey}
\bibinfo{author}{Tan, C.}, \bibinfo{author}{Sun, F.}, \bibinfo{author}{Kong, T.}, \bibinfo{author}{Zhang, W.}, \bibinfo{author}{Yang, C.}, \bibinfo{author}{Liu, C.}, \bibinfo{year}{2018}.
\newblock \bibinfo{title}{A survey on deep transfer learning}.
\newblock \bibinfo{journal}{Artificial Intelligence Review} \bibinfo{volume}{52}, \bibinfo{pages}{1--40}.
\bibitem[{Tasinkevych(2019)}]{andrewmvd2019isic}
\bibinfo{author}{Tasinkevych, A.}, \bibinfo{year}{2019}.
\newblock \bibinfo{title}{{ISIC} 2019: Skin lesion analysis towards melanoma detection}.
\newblock \bibinfo{howpublished}{\url{https://www.kaggle.com/datasets/andrewmvd/isic-2019}}.
\newblock \bibinfo{note}{Accessed: 2023-03-01}.
\bibitem[{Team(2023)}]{ImageNet_2023}
\bibinfo{author}{Team, I.}, \bibinfo{year}{2023}.
\newblock \bibinfo{title}{Imagenet}.
\newblock \URLprefix \url{https://www.image-net.org/}. \bibinfo{note}{online; accessed 12-April-2023}.
\bibitem[{Thapa and Camtepe(2021)}]{thapa2021precision}
\bibinfo{author}{Thapa, C.}, \bibinfo{author}{Camtepe, S.}, \bibinfo{year}{2021}.
\newblock \bibinfo{title}{Precision health data: Requirements, challenges and existing techniques for data security and privacy}.
\newblock \bibinfo{journal}{Computers in biology and medicine} \bibinfo{volume}{129}, \bibinfo{pages}{104130}.
\bibitem[{Turgutlu(2022)}]{turgutlu_self_supervised}
\bibinfo{author}{Turgutlu, K.}, \bibinfo{year}{2022}.
\newblock \bibinfo{title}{Self-supervised learning library}.
\newblock \URLprefix \url{https://github.com/KeremTurgutlu/self_supervised}. \bibinfo{note}{available at: \url{https://github.com/KeremTurgutlu/self_supervised}}.
\bibitem[{Vaswani et~al.(2017)Vaswani, Shazeer, Parmar, Uszkoreit, Jones, Gomez, Kaiser and Polosukhin}]{vaswani2017attention}
\bibinfo{author}{Vaswani, A.}, \bibinfo{author}{Shazeer, N.}, \bibinfo{author}{Parmar, N.}, \bibinfo{author}{Uszkoreit, J.}, \bibinfo{author}{Jones, L.}, \bibinfo{author}{Gomez, A.N.}, \bibinfo{author}{Kaiser, L.}, \bibinfo{author}{Polosukhin, I.}, \bibinfo{year}{2017}.
\newblock \bibinfo{title}{Attention is all you need}, in: \bibinfo{booktitle}{Advances in Neural Information Processing Systems}, pp. \bibinfo{pages}{5998--6008}.
\bibitem[{Wang et~al.(2020)Wang, Dong, Wang and Wang}]{Wang2020}
\bibinfo{author}{Wang, S.}, \bibinfo{author}{Dong, L.}, \bibinfo{author}{Wang, X.}, \bibinfo{author}{Wang, X.}, \bibinfo{year}{2020}.
\newblock \bibinfo{title}{Classification of pathological types of lung cancer from ct images by deep residual neural networks with transfer learning strategy}.
\newblock \bibinfo{journal}{Open Medicine (Warsaw)} \bibinfo{volume}{15}, \bibinfo{pages}{190--197}.
\newblock \DOIprefix\doi{10.1515/med-2020-0028}.
\bibitem[{Warnakulasuriya(2009)}]{warnakulasuriya2009global}
\bibinfo{author}{Warnakulasuriya, S.}, \bibinfo{year}{2009}.
\newblock \bibinfo{title}{Global epidemiology of oral and oropharyngeal cancer}.
\newblock \bibinfo{journal}{Oral Oncology} \bibinfo{volume}{45}, \bibinfo{pages}{309--316}.
\bibitem[{Wu et~al.(2019)Wu, Shen and Van Den~Hengel}]{wu2019wider}
\bibinfo{author}{Wu, Z.}, \bibinfo{author}{Shen, C.}, \bibinfo{author}{Van Den~Hengel, A.}, \bibinfo{year}{2019}.
\newblock \bibinfo{title}{Wider or deeper: Revisiting the resnet model for visual recognition}.
\newblock \bibinfo{journal}{Pattern recognition} \bibinfo{volume}{90}, \bibinfo{pages}{119--133}.
\bibitem[{Yadav et~al.(2019)Yadav, Jain and Shinde}]{Yadav2019}
\bibinfo{author}{Yadav, S.}, \bibinfo{author}{Jain, A.K.}, \bibinfo{author}{Shinde, P.}, \bibinfo{year}{2019}.
\newblock \bibinfo{title}{A survey on deep learning techniques for lung cancer detection}.
\newblock \bibinfo{journal}{International Journal of Innovative Technology and Exploring Engineering (IJITEE)} \bibinfo{volume}{8}, \bibinfo{pages}{1216--1220}.
\newblock \URLprefix \url{https://www.ijitee.org/wp-content/uploads/papers/v8i10s/J10430881019.pdf}.
\bibitem[{You et~al.(2019)You, Long, Wang and Jordan}]{you2019learning}
\bibinfo{author}{You, K.}, \bibinfo{author}{Long, M.}, \bibinfo{author}{Wang, J.}, \bibinfo{author}{Jordan, M.I.}, \bibinfo{year}{2019}.
\newblock \bibinfo{title}{How does learning rate decay help modern neural networks?}
\newblock \bibinfo{journal}{arXiv preprint arXiv:1908.01878} .
\bibitem[{Zbontar et~al.(2021)Zbontar, Jing, Misra, LeCun and Deny}]{zbontar2021barlow}
\bibinfo{author}{Zbontar, J.}, \bibinfo{author}{Jing, L.}, \bibinfo{author}{Misra, I.}, \bibinfo{author}{LeCun, Y.}, \bibinfo{author}{Deny, S.}, \bibinfo{year}{2021}.
\newblock \bibinfo{title}{Barlow twins: Self-supervised learning via redundancy reduction}, in: \bibinfo{booktitle}{International Conference on Learning Representations (ICLR)}.
\bibitem[{Zhang et~al.(2019)Zhang, Xie, Xia and Shen}]{zhang2019attention}
\bibinfo{author}{Zhang, J.}, \bibinfo{author}{Xie, Y.}, \bibinfo{author}{Xia, Y.}, \bibinfo{author}{Shen, C.}, \bibinfo{year}{2019}.
\newblock \bibinfo{title}{Attention residual learning for skin lesion classification}.
\newblock \bibinfo{journal}{IEEE transactions on medical imaging} \bibinfo{volume}{38}, \bibinfo{pages}{2092--2103}.
\bibitem[{Zhuang et~al.(2021)Zhuang, Qi, Duan, Xi, Zhu, Zhu, Xiong and He}]{zhuang2020comprehensive}
\bibinfo{author}{Zhuang, F.}, \bibinfo{author}{Qi, Z.}, \bibinfo{author}{Duan, K.}, \bibinfo{author}{Xi, D.}, \bibinfo{author}{Zhu, Y.}, \bibinfo{author}{Zhu, H.}, \bibinfo{author}{Xiong, H.}, \bibinfo{author}{He, Q.}, \bibinfo{year}{2021}.
\newblock \bibinfo{title}{A comprehensive survey on transfer learning}.
\newblock \bibinfo{journal}{Proceedings of the IEEE} \bibinfo{volume}{109}, \bibinfo{pages}{43--76}.

\end{thebibliography}
\clearpage
\appendix
\section{Appendix}
\subsection{Fine-tuning implementation details}
\label{sec:fine_tuning_details}
The ISIC2019 dataset consists of images of varying dimensions, but deep learning models such as ResNet-50 require an input of homogeneous dimensionality. Therefore, we resize the data to $256\times256$. A batch size of 64 is used to suit our deep learning models. During training,   we apply random data augmentation before passing a mini-batch through the network where each element of the batch is randomly cropped, rotated and flipped. The rotation is by a random angle in $[0,45]$ degrees, and the resize scale and resize ratio for cropping are $(0.7, 1.0)$ and $(0.75, 1.33)$, respectively. Data augmentation applied in this way during supervised learning in a standard strategy to prevent overfitting. The models observe several slightly different views of each image during training which can be seen in Figure \ref{fig:cancer_linear}. Each column represents a mini-batch of size 2, with data augmentation. This is an example of slightly different views of the same data which is presented to the model during training.


Test time augmentation is another strategy to improve generalisation, closely related to the above technique \cite{9710313}. The standard way of computing probabilities at test time is to take the test data, pass it through the neural network, and return an output probability through a softmax layer. Test time augmentation simply repeats this process several times, for augmented views of the test data. This will return several probabilities for each test input, which are then averaged to get the final probability. For example, in Figure \ref{fig:cancer_linear}, test time augmentation would compute the probabilities for each column the standard way, and then average the probabilities. This is because the columns represent the same data but under different augmentation. We use exactly the same augmentations as during training, i.e. as in Table \ref{trainaugs}, and compute the average across three probabilities for each test input. Predictions are then made by taking the argmax of these probabilities, as usual.

\begin{figure}[htbp!]
    \centering
    \includegraphics[scale=0.45]{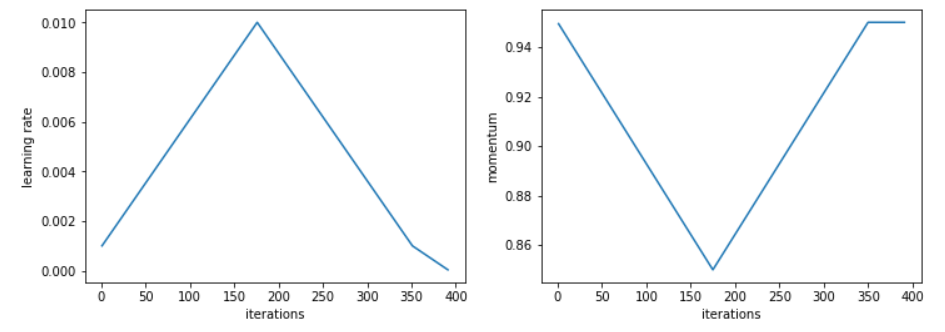}
    \caption{1cycle policy showing the learning rate and momentum follow inverse schedules \cite{fastai_onecycle}.}
    \label{fig:fitonecycle}
\end{figure}

The 1cycle training policy is a learning rate and momentum scheduler \cite{smith2017super,smith2018disciplined}. It involves starting training with a very low learning rate $l_1$ which is then increased up to a maximal learning value $l_2$. Next, the learning rate is slowly decreased to $l_1$, and for the last several epochs, to a much lower final learning rate $\frac{l_1}{K}$, where $K>>1$. In other words, the learning rate has an increasing period, from $l_1$ up to $l_2$, followed by a decreasing period down to $\frac{l_1}{K}$. The momentum is also scheduled, but inversely to the learning rate. Momentum decreases at the start of training, to a minimum, then is increased to a maximum. The scheme is most easily understood pictorially and can be seen in Figure \ref{fig:fitonecycle}.

\begin{figure}[htbp!]
    \centering
    \includegraphics[scale=0.4]{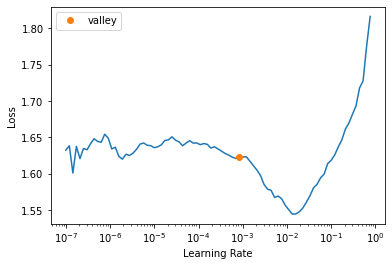}
    \caption{Loss vs. the learning rate.}
    \label{fig:lr_im}
\end{figure}

The 1cycle schedule diverges from standard approaches, which predominantly involve learning rate decay - starting at the maximal learning rate and decreasing to a minimum \cite{you2019learning}. Smith \cite{smith2018disciplined} showed that increasing the learning rate rapidly at the start of training has a regularising effect. It has also been found that when using this policy other forms of regularisation must be reduced - hence we do not consider dropout and similar techniques in this work. Moreover, the policy has a particular advantage when the amount of training data is limited, making it especially suited to our purposes.
An important hyperparameter in the 1cycle policy is the maximal learning rate (i.e. the peak in \ref{fig:fitonecycle}). In fact, this hyperparameter can be automatically inferred. A learning rate finder launches a mock training session and trains the model for several iterations, increasing the learning rate each time and recording the loss. The learning rate starts at a very low value and increases to a high value. A representative plot of this procedure can be seen in \ref{fig:lr_im}. The loss will decrease at the start, before eventually increasing (or perhaps oscillating). A maximal $lr$ is chosen that is somewhere in between a sharp valley and the minimum \cite{fastai_lr_find}. 

\begin{table}[htbp!]
\centering
\begin{tabular}{lccc}
\toprule
 & \text{Crop} & \text{Flip} & \text{Rotate} \\
\midrule
\text{Probability} & 1.0 & 0.25 & 0.25 \\
\bottomrule
\end{tabular}
\caption{Probabilities for augmentations.}
\label{trainaugs}
\end{table}

\subsection{Barlow Twins implementation details}
\label{sec:bt_implementation}


When further pre-training with Barlow Twins, we applied the data augmentation strategy with the same probabilities and order as described in the literature \cite{zbontar2021barlow,grill2020bootstrap} and depicted in Table \ref{tab:bt_aug_prob}. We generally maintained the same level of data augmentation procedures, such as blur intensity and color jitter, with the following exceptions:

\begin{itemize}
\item Solarisation: We used the defaults from the \textit{Kornia} library, which allows greater flexibility, whereas Zbontar et al. \cite{zbontar2021barlow} used the default from \textit{PIL.ImageOps}. Specifically, we used a solarisation threshold ($\textrm{sol\_t}$) of 0.1 and an addition value ($\textrm{sol\_a}$) of 0.1. In our implementation, pixels with values above the threshold are inverted, and then the addition value is added to the affected pixels. In contrast, the implementation by Zbontar et al. uses \texttt{ImageOps.solarize()} without arguments, which implicitly uses a threshold of 128 (equivalent to 0.5 in the normalized [0,1] range) and does not apply an addition step. 

\item Batch-wise augmentation: Several augmentations had a `same\_on\_batch' parameter set to `False', allowing for variation across the batch (e.g., varying blur intensity for each image in a batch). This is a minor improvement over other implementations which will add the same blur intensity to all elements of the batch.
\item Watermark handling: Since the open source DermNet images contain watermarks, we implemented a random dropout augmentation (cutout) to cover the text. This was applied as the second augmentation after cropping, only in the case DermNet is included in the pre-training data.\footnote{This was implemented using a custom RandomCenterDropout class. It randomly drops out (sets to zero) a rectangular region in the center of the image with a 33\% probability. The region's dimensions vary randomly: width from 50 to 100 pixels and height from 185 to 190 pixels, designed to approximately cover the watermark's variable position.}
\end{itemize}

Due to memory constraints, we reduced the batch size from 1024 (used when training on ImageNet) to 128 \cite{zbontar2021barlow}. We utilised PyTorch and FastAI frameworks, along with the self\_supervised library \cite{turgutlu_self_supervised}. Our complete code implementation is available on GitHub\footnote{\url{https://github.com/hamish-haggerty/base-rbt}}.

\begin{center}
\begin{tabular}{lcc}
\hline
Augmentation & $T$ & $T'$ \\
\hline
Random crop probability & 1.0 & 1.0 \\
Cutout probability * & 0.33 & 0.33 \\
Horizontal flip probability & 0.5 & 0.5 \\
Color jittering probability & 0.8 & 0.8 \\
Grayscale conversion probability & 0.2 & 0.2 \\
Gaussian blurring probability & 1.0 & 0.1 \\
Solarization probability & 0.0 & 0.2 \\
\hline
\end{tabular}
\end{center}
\captionof{table}{Probabilities of image augmentations for Barlow Twins training. Cutout is only applied when pre-training on IUD dataset (which contains DermNet).}
\label{tab:bt_aug_prob}


\clearpage
\subsection{Statistical significance}
In this section we display statistical significance tables for fine-tuning results from Table \ref{tab:combined_isic_semi_supervised_results} and linear probe results from Table \ref{tab:combined_isic_linear_probe_results}.

\begin{table}[httb]
\centering
\begin{tabular}{llllccc}
\toprule
\multicolumn{2}{c}{\textbf{Model 1}} & \multicolumn{2}{c}{\textbf{Model 2}} & \textbf{Null} & \multicolumn{2}{c}{\textbf{p-value}} \\
\cmidrule(lr){1-2} \cmidrule(lr){3-4} \cmidrule(lr){6-7}
\textbf{Initial} & \textbf{Further} & \textbf{Initial} & \textbf{Further} & \textbf{hypothesis} & \textbf{Acc} & \textbf{F1} \\
\textbf{Method} & \textbf{Method} & \textbf{Method} & \textbf{Method} & & & \\
\midrule
SSL & - & SL & - & \multirow{4}{*}{$\leq$} & $<10^{-10}$ & $<10^{-10}$ \\
SSL & SSL & SSL & - &  & 0.00147 & $<10^{-6}$ \\
SSL & SSL$_p$ & SSL & - &  & 0.00796 & $<10^{-4}$ \\
SL & SSL & SL & - &  & $<10^{-19}$ & $<10^{-13}$ \\
SL & SSL$_p$ & SL & - &  & $<10^{-13}$ & $<10^{-14}$ \\
\midrule
SSL & SSL & SSL & SSL$_p$ & \multirow{3}{*}{=} & 0.60469 & 0.78867 \\
SL & SSL & SL & SSL$_p$ &  & 0.47254 & 0.27684 \\
SL & SSL$_p$ & SSL & SSL$_p$ &  & 0.79869 & 0.95218 \\
\bottomrule
\end{tabular}
\caption{Statistical significance of fine-tuning results from Table \ref{tab:isic_semi_sup_sig} featuring: 1) comparison of models pre-trained once, (row 1); 2) comparison between models pre-trained once and pre-trained twice (rows 2-5); and 3) comparison between SSL and SSL$_p$ for further pre-training (final 3 rows). The null hypothesis is Model 1 $\leq$ Model 2 or Model 1 $=$ Model 2, as indicated. We use the IUD dataset for all further pre-training  and report p-values for mean accuracy and mean F1 score.}
\label{tab:isic_semi_sup_sig}
\vspace{5em}
\centering
\small
\begin{tabular}{llllllcc}
\toprule
\multicolumn{3}{c}{\textbf{Model 1}} & \multicolumn{3}{c}{\textbf{Model 2}} & \multicolumn{2}{c}{\textbf{p-value}} \\
\cmidrule(lr){1-3} \cmidrule(lr){4-6} \cmidrule(lr){7-8}
\textbf{Initial} & \textbf{Further} & \textbf{Further} & \textbf{Initial} & \textbf{Further} & \textbf{Further} & \textbf{Acc} & \textbf{F1} \\
\textbf{Pre-train} & \textbf{Dataset} & \textbf{Method} & \textbf{Pre-train} & \textbf{Dataset} & \textbf{Method} & & \\
\midrule
SSL & - & - & SL & - & - & $<10^{-37}$ & $<10^{-37}$ \\
SL & IUD & SSL$_p$ & SL & - & - & $<10^{-39}$ & $<10^{-42}$ \\
\bottomrule
\end{tabular}
\caption{Statistical significance of linear probe results from Table \ref{tab:combined_isic_linear_probe_results}. The null hypothesis for each row is Model 1 $\leq$ Model 2. p-values are for accuracy (Acc) and F1 score.}
\label{tab:isic_linear_probe_sig}
\end{table}

\vskip3pt 
\end{document}